\documentclass[apj]{emulateapj}
\newcommand{\etal}{\textit{et~al.}\/}
\newcommand{\ie}{\textit{i.e.}\/}
\newcommand{\eg}{\textit{e.g.}\/}
\newcommand{\cf}{\textit{\it cf.}\/}

\shorttitle{Formaldehyde Densitometry of Starburst Galaxies}
\shortauthors{Mangum \etal}

\begin{document}
\title{Formaldehyde Densitometry of Starburst Galaxies}

\author{Jeffrey G.~Mangum}
\affil{National Radio Astronomy Observatory, 520 Edgemont Road,
  Charlottesville, VA  22903, USA}
\email{jmangum@nrao.edu}

\author{Jeremy Darling}
\affil{Center for Astrophysics and Space Astronomy, Department of Astrophysical
and Planetary Sciences, Box 389, University of Colorado, Boulder, CO 
80309-0389, USA}
\email{jdarling@origins.colorado.edu}

\author{Karl M.~Menten}
\affil{Max Planck Instit\"ut f\"ur Radioastronomie, Auf dem H\"ugel
  69, 53121 Bonn, Germany}
\email{kmenten@mpifr-bonn.mpg.de}

\and

\author{Christian Henkel}
\affil{Max Planck Instit\"ut f\"ur Radioastronomie, Auf dem H\"ugel
  69, 53121 Bonn, Germany}
\email{chenkel@mpifr-bonn.mpg.de}

\begin{abstract}

With a goal toward deriving the physical conditions in external
galaxies, we present a survey of the formaldehyde emission in a sample
of starburst systems.  By extending a technique used to derive the
spatial density in star formation regions in our own Galaxy, we show
how the relative intensity of the $1_{10}-1_{11}$ and $2_{11}-2_{12}$
K-doublet transitions of H$_2$CO can provide an accurate
densitometer for the active star formation environments found in
starburst galaxies.  Relying upon an assumed kinetic temperature
and co-spatial emission and absorption from both H$_2$CO transitions,
our technique is applied to a sample of 
nineteen IR-bright galaxies which exhibit various forms of starburst
activity.  In the five galaxies of our sample where both H$_2$CO
transitions were detected we have derived spatial densities.
We also use H$_2$CO to estimate the dense gas mass in our starburst
galaxy sample, finding similar mass estimates for the dense gas
forming stars in these objects as derived using other dense gas
tracers.  A related trend can be seen when one compares $L_{IR}$ to
our derived $n(H_2)$ for the five galaxies within which we have
derived spatial densities.  Even though our number statistics are
small, there appears to be a trend toward higher spatial density for
galaxies with higher infrared luminosity.  This is likely another
representation of the $L_{IR}$-$M_{dense}$ correlation.
\end{abstract}


\section{Introduction}
\label{intro}

Studies of the distribution of Carbon Monoxide (CO) emission in
external galaxies (\cf\ \cite{Young1991}) have pointed to the presence
of large quantities of molecular material in these systems.  These
studies have yielded 
a detailed picture of the molecular mass in many external galaxies.
But, because emission from the abundant CO molecule is generally
dominated by 
radiative transfer effects, such as high optical depth, it is not a
reliable monitor of the physical conditions, such as spatial density
and kinetic temperature, quantities necessary to assess the
possibility of star formation.  Emission from 
less-abundant, higher-dipole moment molecules 
are better-suited to the task of deriving the spatial density and
kinetic temperature of the dense gas in our and external galaxies.  For this
reason, emission line studies from a variety of molecules have been
made toward mainly nearby galaxies (see \cite{Mauersberger1989} (CS),
\cite{Gao2004a} (HCN), 
\cite{Nguyen1992} (HCO$^+$), \cite{Mauersberger1990} and
\cite{Meier2005} (HC$_3$N),
\cite{Mauersberger2003} (NH$_3$), or \cite{Henkel1991} for a review).

The most extensive sets of measurements of molecular line emission
in external galaxies has been done using the J=1-0 transitions of CO
\citep{Helfer2003} and HCN \citep{Gao2004a}.  Since the J=1-0
transitions of CO and HCN are good tracers of the more generally
distributed and the denser gas, respectively, but do not provide
comprehensive information 
about the individual physical conditions of the dense, potentially
star-forming gas, another molecule must be observed for this purpose. 
Formaldehyde (H$_2$CO) has proven to be a reliable density
and kinetic temperature probe in Galactic molecular clouds.  
Existing measurements of the H$_2$CO $1_{10}-1_{11}$ and
  $2_{11}-2_{12}$ emission in a wide variety of galaxies by
  \cite{Baan1986}, \cite{Baan1990}, \cite{Baan1993}, and
  \cite{Araya2004} have mainly concentrated on measurements of the
  $1_{10}-1_{11}$ transition.  One of our goals with the present study
  was to obtain a uniform set of measurements of both K-doublet
  transitions with which the physical conditions, specifically the
  spatial density, in the extragalactic context could be derived.
Using the
unique density selectivity of the K-doublet transitions of H$_2$CO we
have measured the spatial density in a sample of galaxies exhibiting
starburst phenomena and/or high infrared luminosity.  In
\S\ref{H2coProbe} we discuss the specific properties of the H$_2$CO
molecule which make it a good probe of spatial density.
\S\ref{Observations} presents our observation summary; \S\ref{Results}
our H$_2$CO, OH, H111$\alpha$, and continuum emission measurement
results; \S\ref{Analysis} analyses of our H$_2$CO, OH, and
H111$\alpha$ measurements, including Large Velocity Gradient (LVG)
model fits to and dense gas mass calculations based on our H$_2$CO
measurements.

\section{Formaldehyde as a High Density Probe}
\label{H2coProbe}

\begin{figure}
\resizebox{\hsize}{!}{
\includegraphics[scale=0.65]{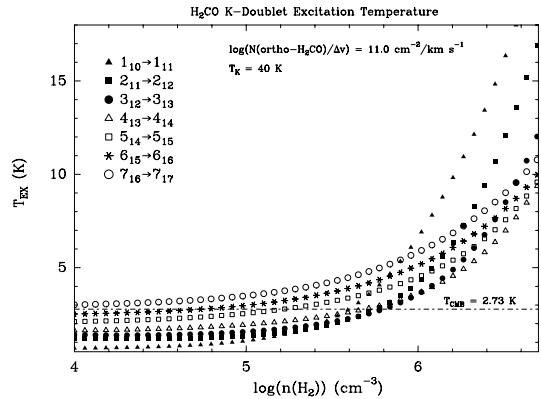}} 
\caption{H$_2$CO K-doublet excitation temperatures as a function of
  molecular hydrogen density at an ortho-H$_2$CO column density per
  velocity gradient of $10^{11.0}$ cm$^{-2}$/km s$^{-1}$ and kinetic
  temperature of 40 K.  Note how
  the J$\lesssim$5 K-doublet transitions go into absorption for
  n(H$_2$) $\lesssim$ 10$^{5.5}$ cm$^{-3}$.}
\label{fig:kdoubtex}
\end{figure}

Formaldehyde (H$_2$CO) is a proven tracer of the high density environs
of molecular clouds.  It is ubiquitous: H$_2$CO is associated with 80\%
of the HII regions surveyed by \cite{Downes1980}, and possesses a
large number of observationally accessible transitions from centimeter
to far-infrared wavelengths.  Because H$_2$CO is a slightly asymmetric
rotor molecule, most rotational energy levels are split by this
asymmetry into two energy levels.  Therefore, the energy levels must
be designated by a total angular momentum quantum number, J, the
projection of J along the symmetry axis for a limiting prolate
symmetric top, K$_{-1}$, and the projection of J along the symmetry
axis for a limiting oblate symmetric top, K$_{+1}$.  This splitting
leads to two basic types of transitions: the high-frequency $\Delta$J
= 1, $\Delta$K$_{-1}$ = 0, $\Delta$K$_{+1}$ = $-$1 ``P-branch''
transitions and the lower-frequency $\Delta$J = 0, $\Delta$K$_{-1}$ =
0, $\Delta$K$_{+1}$ = $\pm$1 ``Q-branch'' transitions, popularly known
as the ``K-doublet'' transitions (see discussion in
\cite{Mangum1993}).  The P-branch transitions are only seen in
emission in regions where n($H_2$) $\gtrsim$ 10$^{4}$ cm$^{-3}$.  The
excitation of the K-doublet 
transitions, though, is not so simple.  For n($H_2$) $\lesssim$ 10$^{5.5}$
cm$^{-3}$, the lower energy states of the $1_{10}-1_{11}$
through $5_{14}-5_{15}$ K-doublet transitions become
overpopulated due to a collisional selection effect (\cite{Evans1975};
\cite{Garrison1975}).  This overpopulation cools the J $\leq$ 5
K-doublets to 
excitation temperatures lower than that of the cosmic microwave
background, causing them to appear in absorption.  For n($H_2$)
$\gtrsim$ $10^{5.5}$ cm$^{-3}$, this collisional pump is quenched and the
J $\leq$ 5 K-doublets are then seen in emission over a wide range of
kinetic temperatures and abundances (see Figure~\ref{fig:kdoubtex}).

\section{Observations}
\label{Observations}

The measurements reported here were made using the National Radio
Astronomy Observatory (NRAO\footnote{The National Radio Astronomy
  Observatory is a facility of the National Science Foundation
  operated under cooperative agreement by Associated Universities,
  Inc.}) Green Bank Telescope (GBT) during the 
period 2006/06/07-14.  Single-pointing measurements were obtained of the 
$1_{10}-1_{11}$ (4.829660 GHz) and $2_{11}-2_{12}$
(14.488479 GHz) K-doublet transitions of H$_2$CO, the 
H111$\alpha$ radio recombination line (RRL) at 4.744183 GHz, and two of
the three ($F=1-0$ and $1-1$) hyperfine satellite lines of the
rotationally excited $^2\Pi_{1/2} J=1/2$ state of OH at 4.750656(3) and
4.765562(3)~GHz \citep{Radford68,Lovas1992} toward a sample of
19 infrared-luminous and/or starburst 
galaxies (Table~\ref{tab:sources}).  Our source sample was chosen to
represent both galaxies with measured H$_2$CO emission or absorption
(\cite{Baan1993}, \cite{Araya2004}) and other galaxies with substantial
molecular emission deduced from HCN measurements (\cite{Gao2004a}).
The single-beam H$_2$CO 
$1_{10}-1_{11}$, H111$\alpha$, and OH $^2\Pi_{1/2} J=1/2$
transitions were measured using 4 spectral windows each with 50 MHz of
bandwidth sampled by 16384 channels.  All of the 4.8 GHz measurements
($\theta_{B} = 153^{\prime\prime}$) utilized the position switching 
technique with reference position located 30 arcmin west in azimuth
from each source position.  The H$_2$CO $2_{11}-2_{12}$ transition was
measured using the dual-beam ($\theta_{B} = 51^{\prime\prime}$; beam
separation 330$^{\prime\prime}$ in 
cross-elevation) receiver over 50 MHz of bandwidth sampled by 16384
channels.  The dual-beam system at 14.5 GHz allowed for both position
switching and beam nodding.  Both correlator configurations produced
a spectral channel width of 3.052 kHz, which is approximately 0.2 and
0.08 km~s$^{-1}$ at 4.8 and 14.5 GHz, respectively.

\begin{deluxetable*}{lcccccl} 
\tabletypesize{\scriptsize}
\tablewidth{0pt}
\tablecolumns{7}
\tablecaption{Source List}
\tablehead{
\colhead{Source} & 
\colhead{$\alpha$} & 
\colhead{$\delta$} & 
\colhead{v$_{hel}$\tablenotemark{a}} & 
\colhead{D$_L$\tablenotemark{b}} &
\colhead{T$_{dust}$\tablenotemark{c}} &
\colhead{Classification\tablenotemark{d}} \\
& \colhead{(J2000)} &\colhead{(J2000)}& 
\colhead{(km s$^{-1}$)} & \colhead{(Mpc)} & \colhead{(K)} &
}
\startdata
NGC~253       & 00:47:33.1  & $-$25:17:18 & 251    & 2.5   & 34 & SAB(s)c \\
NGC~520       & 01:24:35.3  & $+$03:47:37 & 2281   & 30.4  & 38 & Pec, Pair, Sbrst \\
NGC~660       & 01:43:01.7  & $+$13:38:36 & 856    & 14.0  & 37 & SB(s)a:pec, HII LINER\\
NGC~891       & 02:22:33.4  & $+$42:20:57 & 529    & 10.0  & 28 & SA(s)b?:sp \\
IC~342        & 03:46:49.7  & $+$68:05:45 & 31     & 3.7   & 30 & SAB(rs)cd \\
Arp~55        & 09:15:55.1  & $+$44:19:55 & 11957  & 159.4 & 36 & Pair \\
NGC~2903      & 09:32:10.1  & $+$21:30:02 & 556    & 6.2   & 29 & SB(s)d \\
UGC~5101      & 09:35:51.6  & $+$61:21:11 & 11810  & 157.5 & 36 & Sy1.5, LINER \\
M~82          & 09:55:52.2  & $+$69:40:47 & 203    & 3.5   & 45 & I0,Sbrst \\
NGC~3079      & 10:01:57.8  & $+$55:40:47 & 1150   & 16.2  & 32 & SB(s)c, LINER \\
IR~10173+0828 & 10:19:59.9  & $+$08:13:34 & 14716  & 196.2 & \nodata &
Sbrst \\
NGC~3628      & 11:20:17.2  & $+$13:35:20 & 847    & 9.6   & 30 & Sb:pec:sp \\
IC~860        & 13:15:04.1  & $+$24:37:01 & 3866   & 51.5  & \nodata &
Sa, Sbrst \\
M~83          & 13:37:00.9  & $-$29:51:57 & 518    & 3.7   & 31 & SAB(s)c \\
IR~15107+0724 & 15:13:13.1  & $+$07:13:27 & 3897   & 52.0  & \nodata &
Sbrst \\
Arp~220       & 15:34:57.1  & $+$23:30:11 & 5434   & 72.5  & 44 & Pair, Sbrst \\
NGC~6240      & 16:52:59.0  & $+$02:24:02 & 7339   & 97.9  & 41 & I0:pec, LINER, Sy2 \\
IR~17468+1320 & 17:49:06.7  & $+$13:19:54 & 4881   & 65.1  & \nodata &
Sbrst \\
NGC~6946      & 20:34:52.3  & $+$60:09:14 & 48     & 5.5   & 30 & SAB(rs)cd \\
\enddata
\tablenotetext{a}{Heliocentric velocity drawn from the
  literature.}
\tablenotetext{b}{Source distance drawn from the literature 
(${\rm v}< 2000$~km~s$^{-1}$) or calculated assuming 
H$_0=75$~km~s$^{-1}$~Mpc$^{-1}$.} 
\tablenotetext{c}{From \cite{Gao2004b}.}
\tablenotetext{d}{From NED\footnote{The NASA/IPAC Extragalactic 
Database (NED) is operated by the Jet Propulsion Laboratory, California 
Institute of Technology, under contract with the National Aeronautics and 
Space Administration.}, Sbrst = starburst galaxy.}
\label{tab:sources}
\end{deluxetable*} 

To calibrate the intensity scale of our measurements, several
corrections need to be considered:
\begin{description}
\item[Opacity:] Historical opacity estimates based on atmospheric
  model calculations using ambient pressure, temperature, and relative
  humidity measurements indicated that $\tau$ at 4.8 and 14.5 GHz were
  $\sim$ 0.01 and 0.02 during our observations (assuming elevation
  $\gtrsim$ 30 degrees).  The opacity corrections
  $\exp(\tau_0\csc(EL))$ are $\lesssim 1.02$ and $\lesssim 1.04$,
  respectively, and are considered negligible in our analysis.
\item[Flux:] Assuming point-source emission, one can use the
  current relation (derived from measurements) for the aperture
  efficiency $\eta_A = 0.71 \exp\left(-\left(0.0163
      \nu(GHz)\right)^2\right)$ to convert antenna temperature to flux
  density.  At 4.8 and 14.5 GHz this yields $\eta_A$ = 0.71 and 0.67,
  respectively. For elevation 90 degrees and zero atmospheric opacity,
  T$_A$/S = $2.846\eta_A$ = 2.0 and 1.9 for 4.8 and 14.5 GHz,
  respectively. 
These are the K/Jy calibration factors used to convert
  our spectra to flux assuming point-source emission.  Note also that
  since the opacity correction is small, $T^*_A = T_A\exp(A\tau_0)
  \simeq \frac{T_A}{\eta_l} \simeq T_A$, where $\eta_l = 0.99$ for the
  GBT.  Using $\eta_{mb} \simeq 1.32 \eta_A$, we can write the main
  beam brightness temperature as $T_{mb} \simeq
  \frac{T^*_A}{\eta_{mb}}$.
\item[Source Structure:] Several of our sources are known to have
  high-density structure (measured with high-dipole moment molecules
  like HCN) on scales approaching the size of our beam at 14.5 GHz
  (51$^{\prime\prime}$).  The beam coupling correction necessary to
  account for structure in our 14.5 GHz measurements, relative to our
  point-source assumption, is given by:
\begin{equation}
  f_{coupling} = \frac{\theta^2_B + \theta^2_S}{\theta^2_B}
\label{eq:fcoupling}
\end{equation}
  This correction factor is less than 20\% for $\theta_S \leq
  23^{\prime\prime}$.  With the exception of M~82, none of the galaxies
  in our sample have measured dense molecular gas structure larger
  than $\sim 20^{\prime\prime}$.  We therefore assume that, with the
  exception of M~82, all of the H$_2$CO emission reported in these
  measurements is from structures smaller than the primary beams of
  our measurements.  This assumption allows us to equate source main
  beam (T$_{mb}$) and radiation (T$_R$) temperature to our antenna
  temperature (T$^*_A$) measurements.
\item[Absolute Amplitude Calibration:] The GBT absolute amplitude
  calibration is reported to be 10-15\% at all frequencies, limited
  mainly by temporal drifts in the noise diodes used as absolute
  amplitude calibration standards.  Relative calibration between our
  4.8 and 14.5 GHz measurements is expected to be much better;
  estimated at $\sim 5\%$.
\end{description}

\section{Results}
\label{Results}

In the following all spectra have been smoothed to 20 km~s$^{-1}$
(H$_2$CO $1_{10}-1_{11}$, OH, and H111$\alpha$) or 10 km~s$^{-1}$
(H$_2$CO $2_{11}-2_{12}$) to both increase the
signal-to-noise ratio of our measurements and more closely match other
molecular spectral line measurements of these galaxies (\ie\
\cite{Gao2004a}).  For each detection we list the peak intensity, velocity
width (FWZI), and integrated intensity derived from direct
channel-by-channel integration of the line profiles.

\subsection{H$_2$CO}
\label{FormResults}

\begin{figure}
\resizebox{\hsize}{!}{
\includegraphics[scale=0.75]{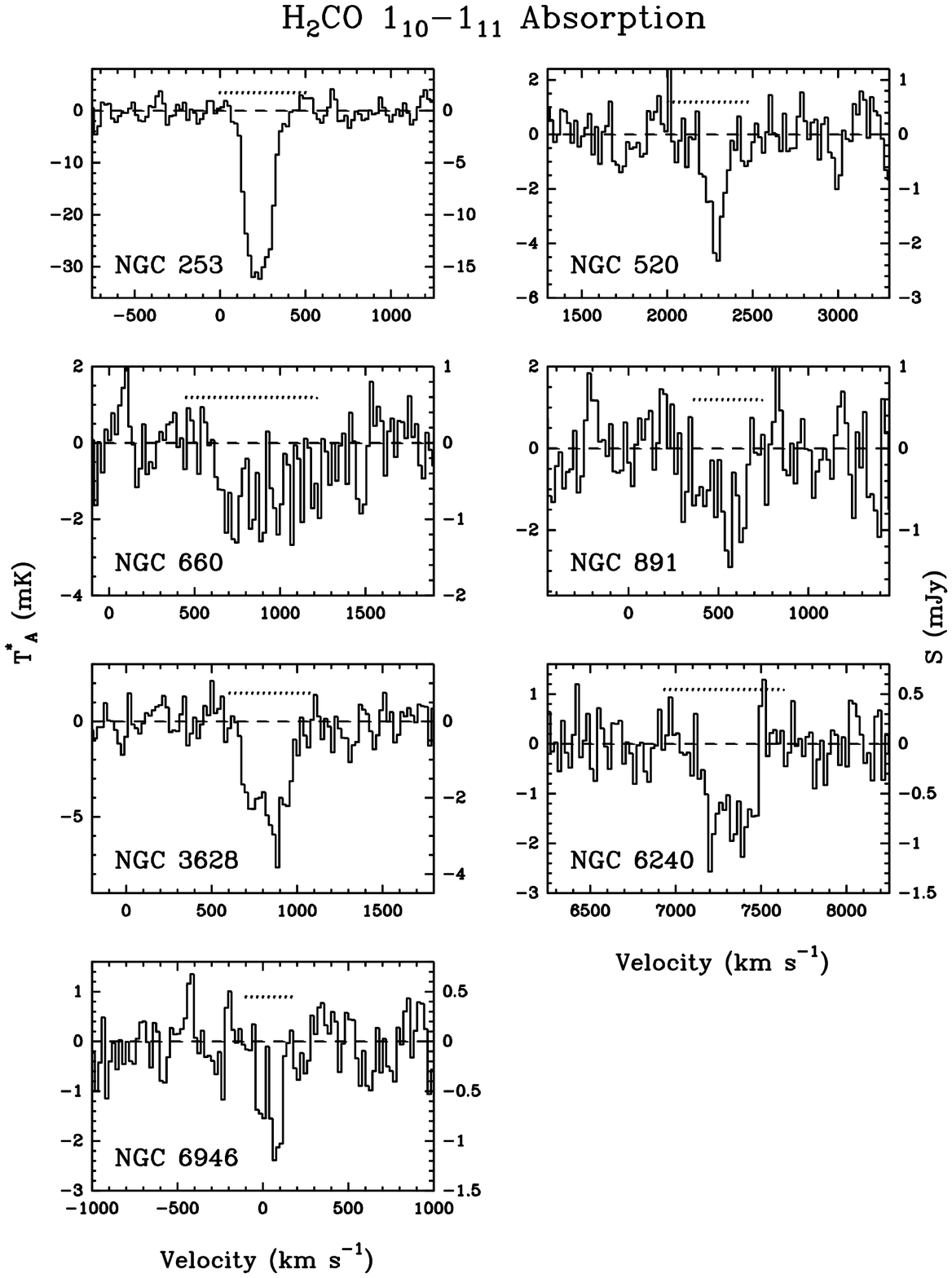}}
\caption{H$_2$CO $1_{10}-1_{11}$ absorption spectra.  The dotted line
  within each spectrum indicates the FWZI CO linewidth.}
\label{fig:ExgalForm6cmAbs}
\end{figure}

\begin{figure}
\resizebox{\hsize}{!}{
\includegraphics[scale=0.75]{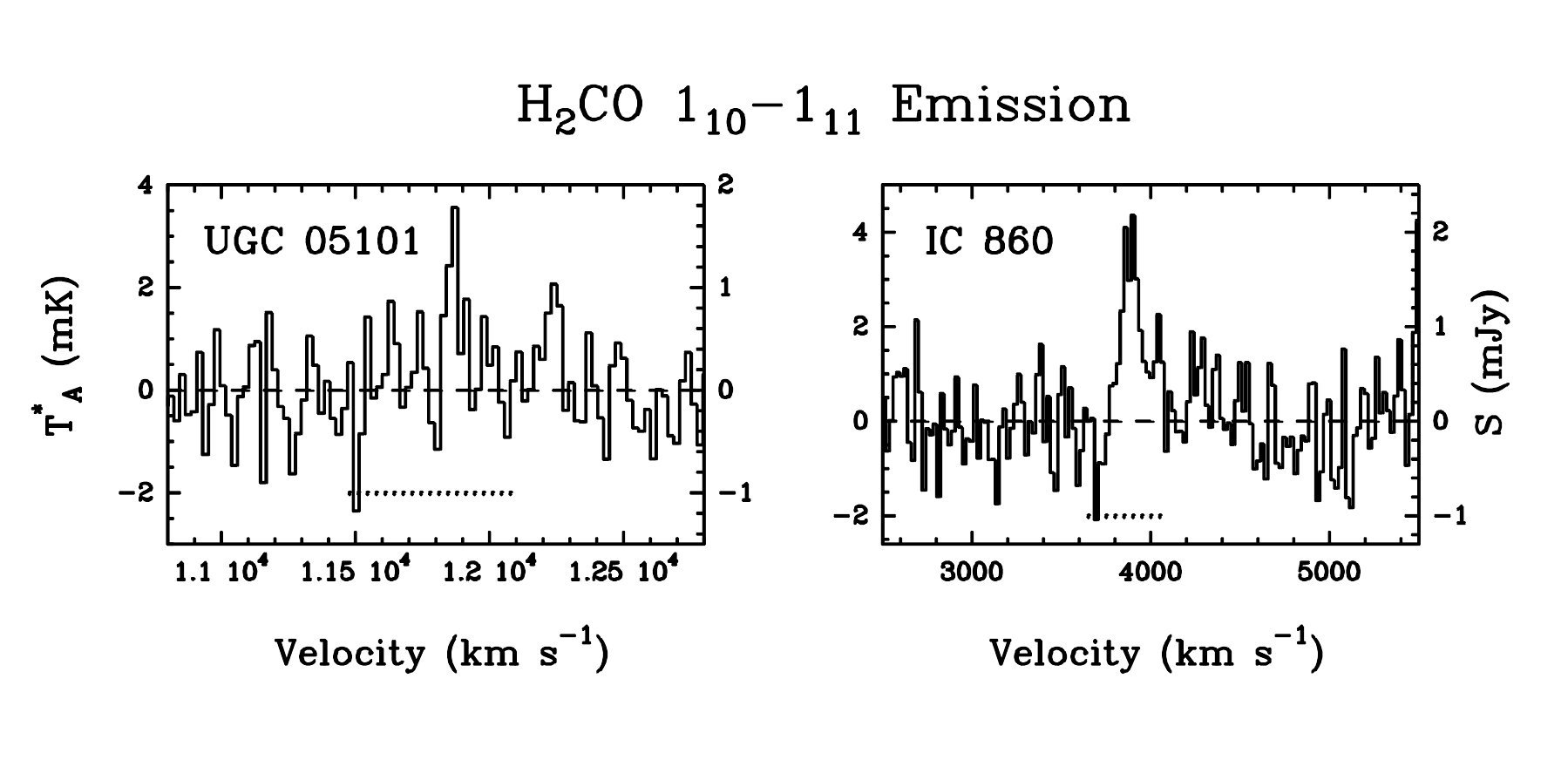}}
\caption{H$_2$CO $1_{10}-1_{11}$ emission spectra.  The dotted line
  within each spectrum indicates the FWZI CO linewidth.}
\label{fig:ExgalForm6cmEmis}
\end{figure}

\begin{figure}
\resizebox{\hsize}{!}{
\includegraphics[scale=0.75]{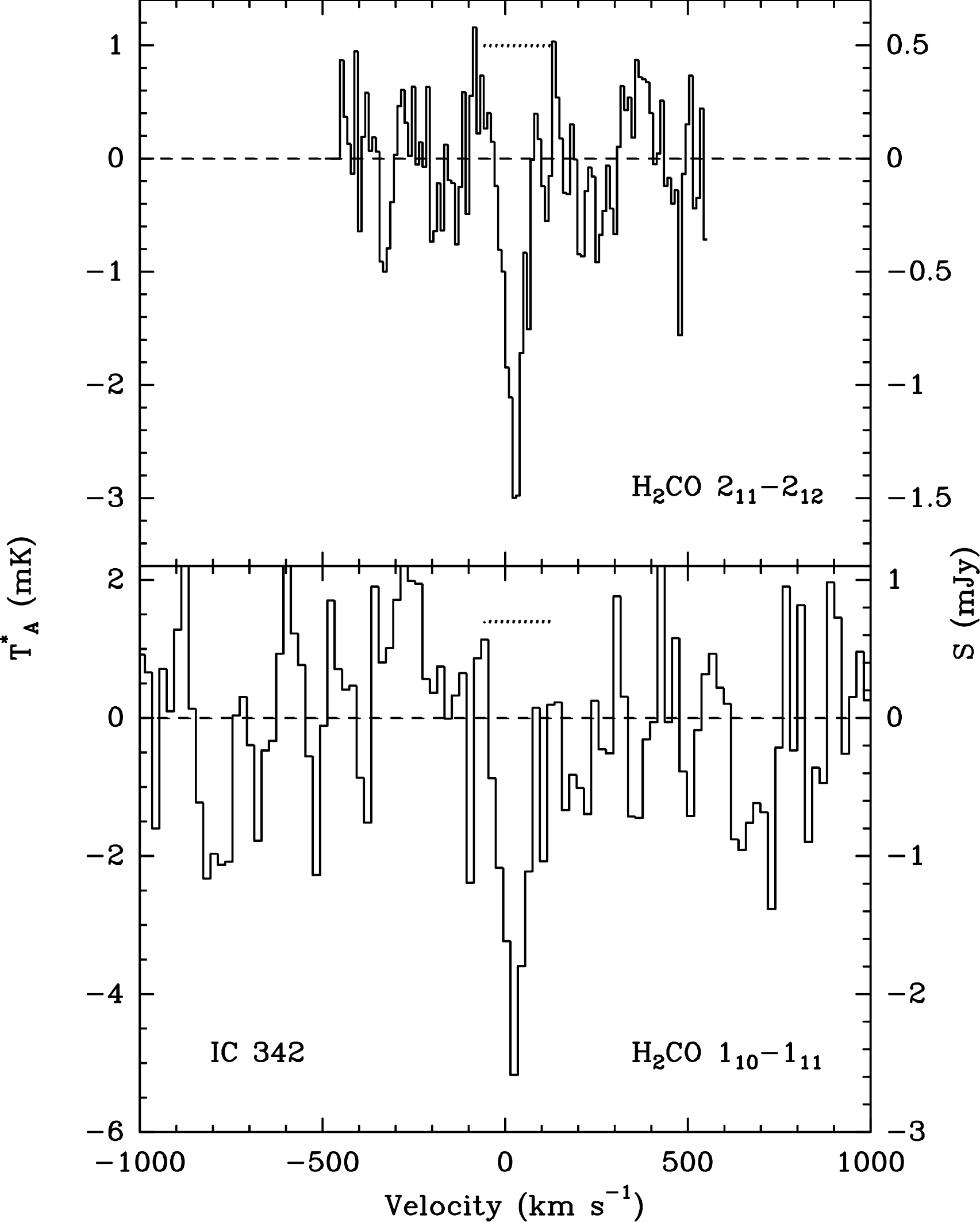}}
\caption{H$_2$CO $2_{11}-2_{12}$ (top) and
  $1_{10}-1_{11}$ (bottom) spectra of IC~342.  The dotted line 
  within each spectrum indicates the FWZI CO linewidth.}
\label{fig:IC342FormSpec}
\end{figure}

\begin{figure}
\resizebox{\hsize}{!}{
\includegraphics[scale=0.75]{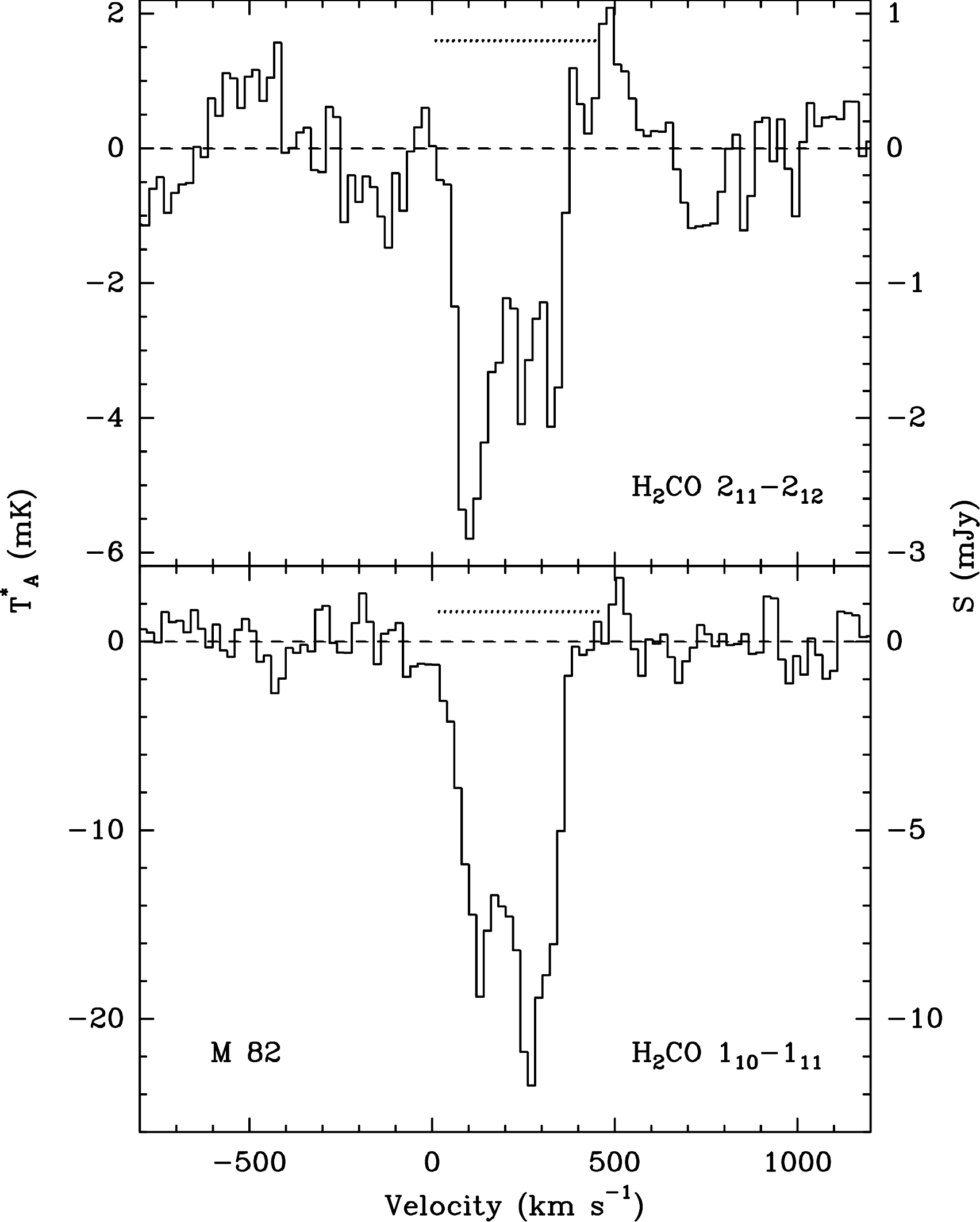}}
\caption{H$_2$CO $2_{11}-2_{12}$ (top) and
  $1_{10}-1_{11}$ (bottom) spectra of M~82.  The dotted line 
  within each spectrum indicates the FWZI CO linewidth.}
\label{fig:M82FormSpec}
\end{figure}

\begin{figure}
\resizebox{\hsize}{!}{
\includegraphics[scale=0.8]{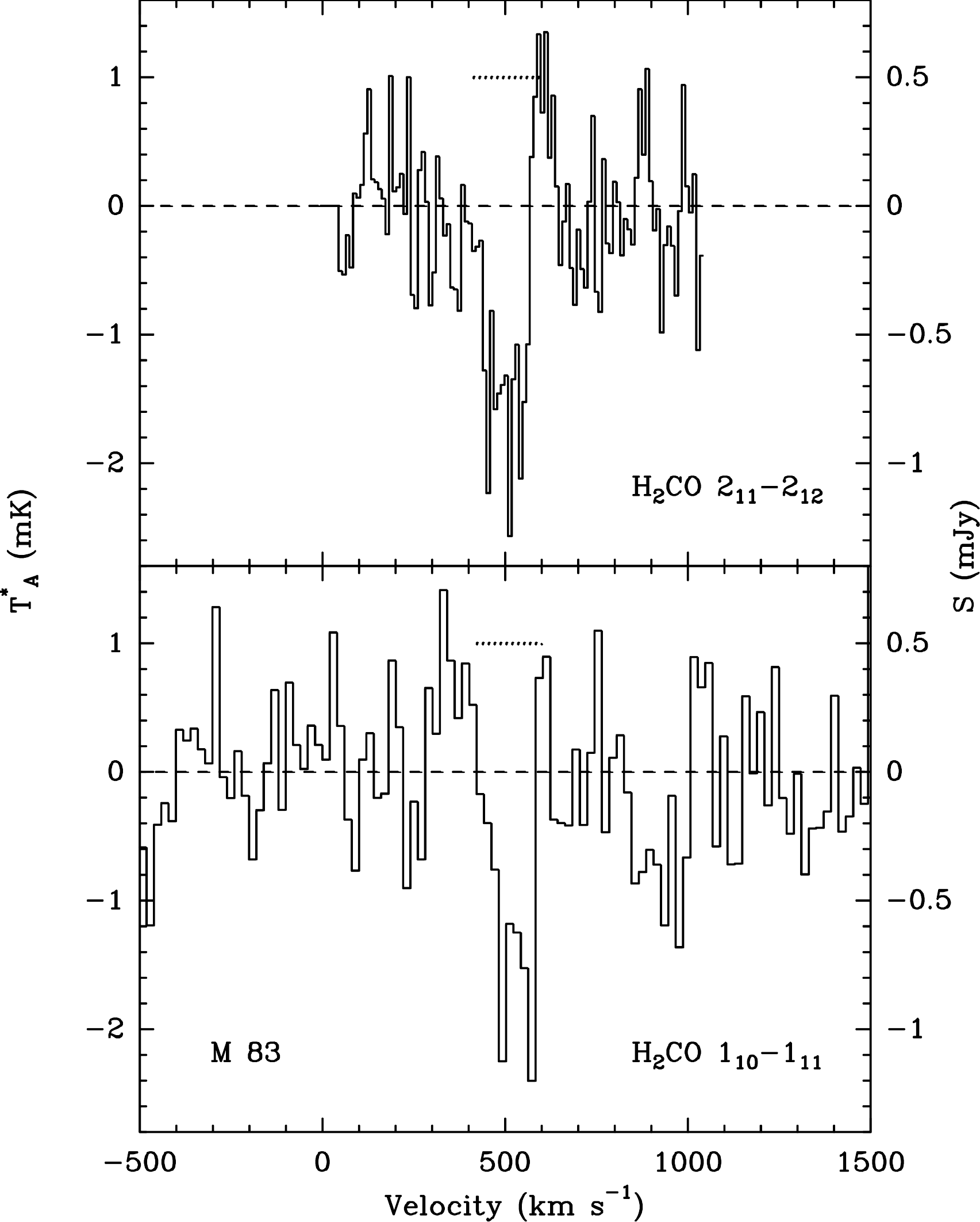}}
\caption{H$_2$CO $2_{11}-2_{12}$ (top) and
  $1_{10}-1_{11}$ (bottom) spectra of M~83.  The dotted line 
  within each spectrum indicates the FWZI CO linewidth.}
\label{fig:M83FormSpec}
\end{figure}

\begin{figure}
\resizebox{\hsize}{!}{
\includegraphics[scale=0.8]{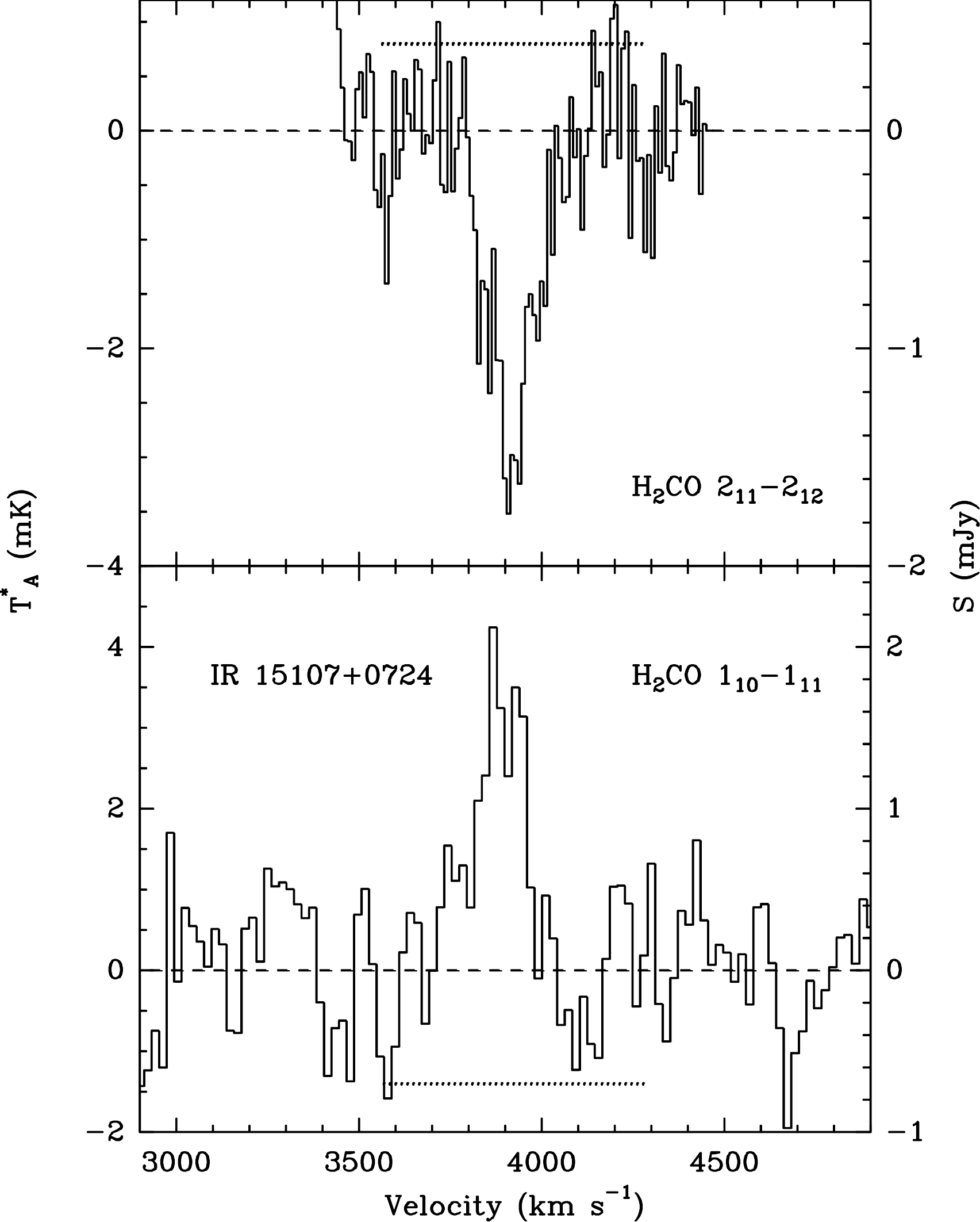}}
\caption{H$_2$CO $2_{11}-2_{12}$ (top) and
  $1_{10}-1_{11}$ (bottom) spectra of IR~15107+0724.  The
  dotted line within each spectrum indicates the FWZI CO linewidth.}
\label{fig:IR15107FormSpec}
\end{figure}

\begin{figure}
\resizebox{\hsize}{!}{
\includegraphics[scale=0.8]{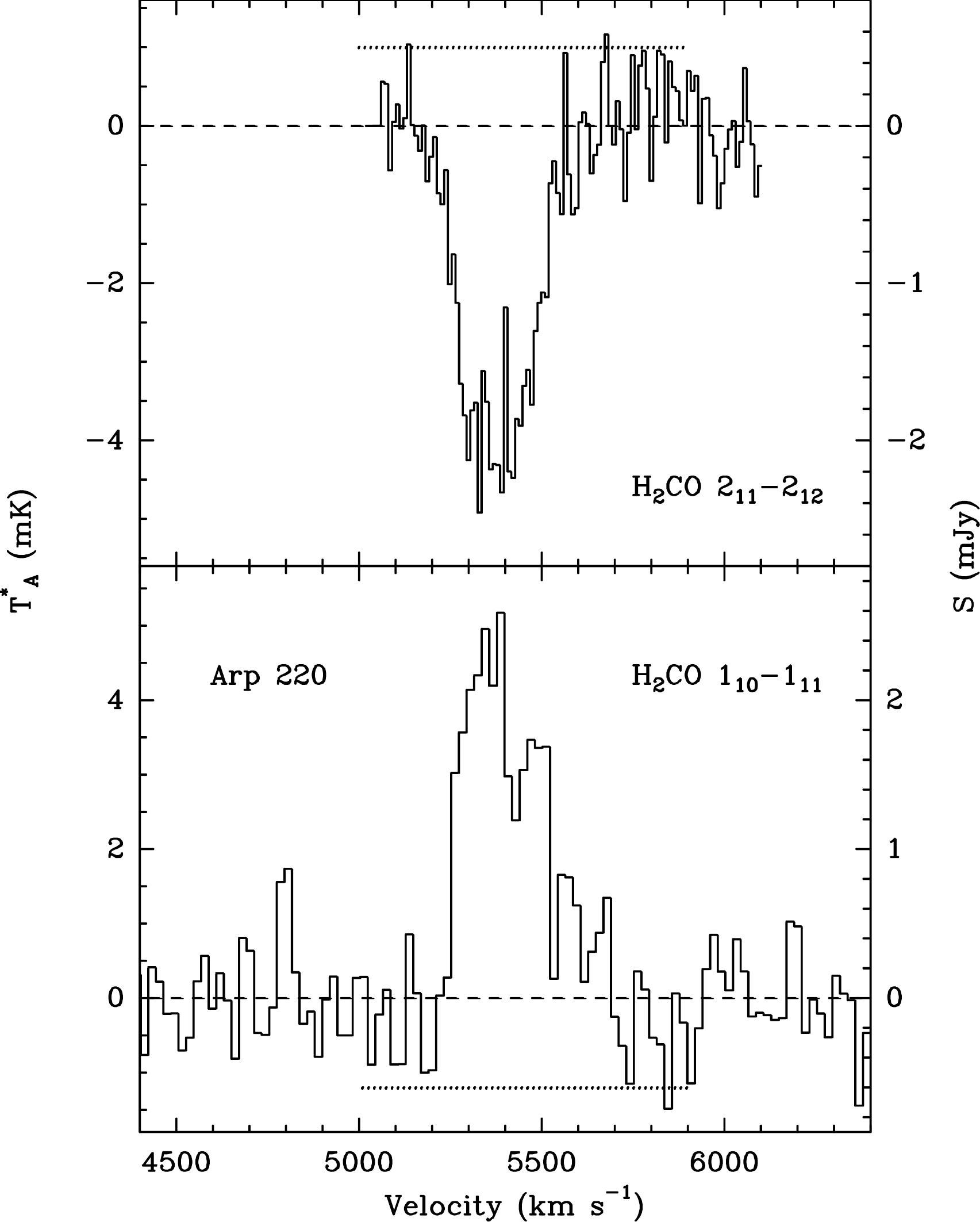}}
\caption{H$_2$CO $2_{11}-2_{12}$ (top) and
  $1_{10}-1_{11}$ (bottom) spectra of Arp~220.  The dotted line 
  within each spectrum indicates the FWZI CO linewidth.}
\label{fig:Arp220FormSpec}
\end{figure}

Table~\ref{tab:h2codetections} lists
our H$_2$CO $1_{10}-1_{11}$ and $2_{11}-2_{12}$ results.
Spectra for the galaxies detected only in $1_{10}-1_{11}$
absorption are displayed in Figure~\ref{fig:ExgalForm6cmAbs}, while
Figure~\ref{fig:ExgalForm6cmEmis} shows spectra from galaxies detected
only in $1_{10}-1_{11}$ emission.  IC~342, M~82, M~83,
IR~15107+0724, and Arp~220 (Figures~\ref{fig:IC342FormSpec},
\ref{fig:M82FormSpec}, 
\ref{fig:M83FormSpec}, \ref{fig:IR15107FormSpec}, and
\ref{fig:Arp220FormSpec}) were all detected in both H$_2$CO
transitions.  

\begin{deluxetable*}{llrrrrr} 
\tabletypesize{\scriptsize}
\tablewidth{0pt}
\tablecolumns{7}
\tablecaption{H$_2$CO Measurements}
\tablehead{
\colhead{Source} & 
\colhead{Transition} & 
\colhead{$T^*_A$/S$_{peak}$\tablenotemark{a}} & 
\colhead{v$_{hel}$\tablenotemark{a}} & 
\colhead{FWZI} & 
\colhead{$\int$T$^*_A$dv/$\int$Sdv\tablenotemark{a}} &
\colhead{$\tau$\tablenotemark{b}} \\
&& \colhead{(mK/mJy)} & 
\colhead{(km s$^{-1}$)} & 
\colhead{(km s$^{-1}$)} & 
\colhead{(mK km s$^{-1}$/mJy km s$^{-1}$)} & \\
}
\startdata
NGC 253       & $1_{10}-1_{11}$ & $-$31.9(1.8)/$-$16.2(0.9) & 226.7(2.2) & 362 &
                $-$5782(165)/$-$2935(84) & 0.0074(4) \\ 
NGC 520       & $1_{10}-1_{11}$ & $-$4.5(0.8)/$-$2.3(0.4) & 2287.6(6.5) & 207 &
                $-$469(61)/$-$238(31)  & 0.0050(9) \\ 
              & $2_{11}-2_{12}$ & (0.52)/(0.29)
              &\nodata&\nodata&\nodata& (0.0007)  \\ 
NGC 660       & $1_{10}-1_{11}$ & $-$2.6(0.8)/$-$1.3(0.4) & 934.6(5.0) & 708 &
                $-$906(87)/$-$460(44) & 0.0025(8) \\
              & $2_{11}-2_{12}$ & (0.47)/(0.26) &\nodata&\nodata&\nodata& (0.0006)  \\
NGC 891       & $1_{10}-1_{11}$ & $-$2.7(0.8)/$-$1.4(0.4) & 499.5(5.0) & 323
              & $-$445(71)/$-$226(36) & 0.0020(6) \\
IC 342        & $1_{10}-1_{11}$ & $-$5.1(1.2)/$-$2.6(0.6) & 25.7(6.9) & 124 &
                $-$317(67)/$-$161(34)  & 0.006(1)* \\ 
              & $2_{11}-2_{12}$ & $-$2.9(0.6)/$-$1.5(0.3) & 26.7(2.8) & 104 &
               $-$140(16)/$-$78(9) & 0.0035(7) \\ 
Arp 55        & $1_{10}-1_{11}$ & (1.80)/(0.90) &\nodata&\nodata&\nodata& (0.002)  \\
NGC 2903      & $1_{10}-1_{11}$ & (0.91)/(0.46) &\nodata&\nodata&\nodata& (0.0009)  \\
UGC 5101      & $1_{10}-1_{11}$ & $+$3.5(0.8)/$+$1.8(0.4) & 11864.6(6.3) & 99 &
               175(51)/89(26) & \nodata \\
M 82          & $1_{10}-1_{11}$ & $-$23.0(1.0)/$-$11.7(0.5) & 214.3(5.0) & 430 &
               $-$4960(110)/$-$2518(56) & 0.0032(1) \\ 
              & $2_{11}-2_{12}$ & $-$5.2(0.7)/$-$2.9(0.4) & 188.8(5.0) & 381 &
               $-$995(56)/$-$553(31) &  0.0015(2)*\\ 
NGC 3079      & $1_{10}-1_{11}$ & (1.30)/(0.66) &\nodata&\nodata&\nodata& (0.0009) \\
IR 10173+0828 & $1_{10}-1_{11}$ & (1.38)/(0.70) &\nodata&\nodata&\nodata& (0.002)  \\
NGC 3628      & $1_{10}-1_{11}$ & $-$7.7(0.8)/$-$3.9(0.4) & 840.8(5.0) & 368 &
               $-$1412(67)/$-$717(34) & 0.0070(7) \\ 
M 83          & $1_{10}-1_{11}$ & $-$2.4(0.6)/$-$1.2(0.3) & 522.9(10.8) & 155 &
               $-$183(42)/$-$93(21) & 0.0017(4)* \\ 
              & $2_{11}-2_{12}$ & $-$2.3(0.4)/$-$1.3(0.2) & 496.8(6.6) & 155 &
               $-$182(18)/$-$101(10) & 0.0026(5) \\ 
IC 860        & $1_{10}-1_{11}$ & $+$4.3(1.0)/$+$2.2(0.5) & 3892.5(8.1) & 267 &
               579(75)/294(38)  & \nodata \\ 
IR 15107+0724 & $1_{10}-1_{11}$ & $+$4.4(0.8)/$+$2.2(0.4) & 3875.3(5.0) & 263 &
               569(73)/289(37)  & \nodata \\ 
              & $2_{11}-2_{12}$ & $-$3.2(0.5)/$-$1.8(0.3) & 3918.4(5.3) & 244 & 
               $-$396(25)/$-$220(14)  & 0.0041(6) \\ 
Arp 220       & $1_{10}-1_{11}$ & $+$5.1(0.6)/$+$2.6(0.3) & 5412.0(5.0) & 447 &
               1145(65)/581(33)  & \nodata \\ 
              & $2_{11}-2_{12}$ & $-$4.5(0.5)/$-$2.5(0.3) & 5379.2(4.6) & 410 &
               $-$923(34)/$-$513(19) & 0.0050(6) \\ 
NGC 6240      & $1_{10}-1_{11}$ & $-$2.6(0.4)/$-$1.3(0.2) & 7316.9(5.0) & 392 &
               $-$502(47)/$-$255(24) & 0.0026(4) \\
IR 17468+1320 & $1_{10}-1_{11}$ & (0.96)/(0.49) &\nodata&\nodata&\nodata& (0.003)*  \\
NGC 6946      & $1_{10}-1_{11}$ & $-$2.4(0.6)/$-$1.2(0.3) & 49.8(5.0) & 210 &
               $-$248(35)/$-$126(18) & 0.0022(6) \\ 
              & $2_{11}-2_{12}$ & (0.43)/(0.24) &\nodata&\nodata&\nodata& (0.0005)  \\
\enddata
\tablenotetext{a}{~Heliocentric optical velocity frame.  RMS noise
  levels are for 20 ($J=1$) and 10 ($J=2$) km~s$^{-1}$ channels.} 
\tablenotetext{b}{~See \S\ref{OpticalDepth} for further information.}
\label{tab:h2codetections}
\end{deluxetable*} 

\subsection{OH and H111$\alpha$}
\label{OHH111Results}

\begin{figure}
\resizebox{\hsize}{!}{
\includegraphics[scale=0.75]{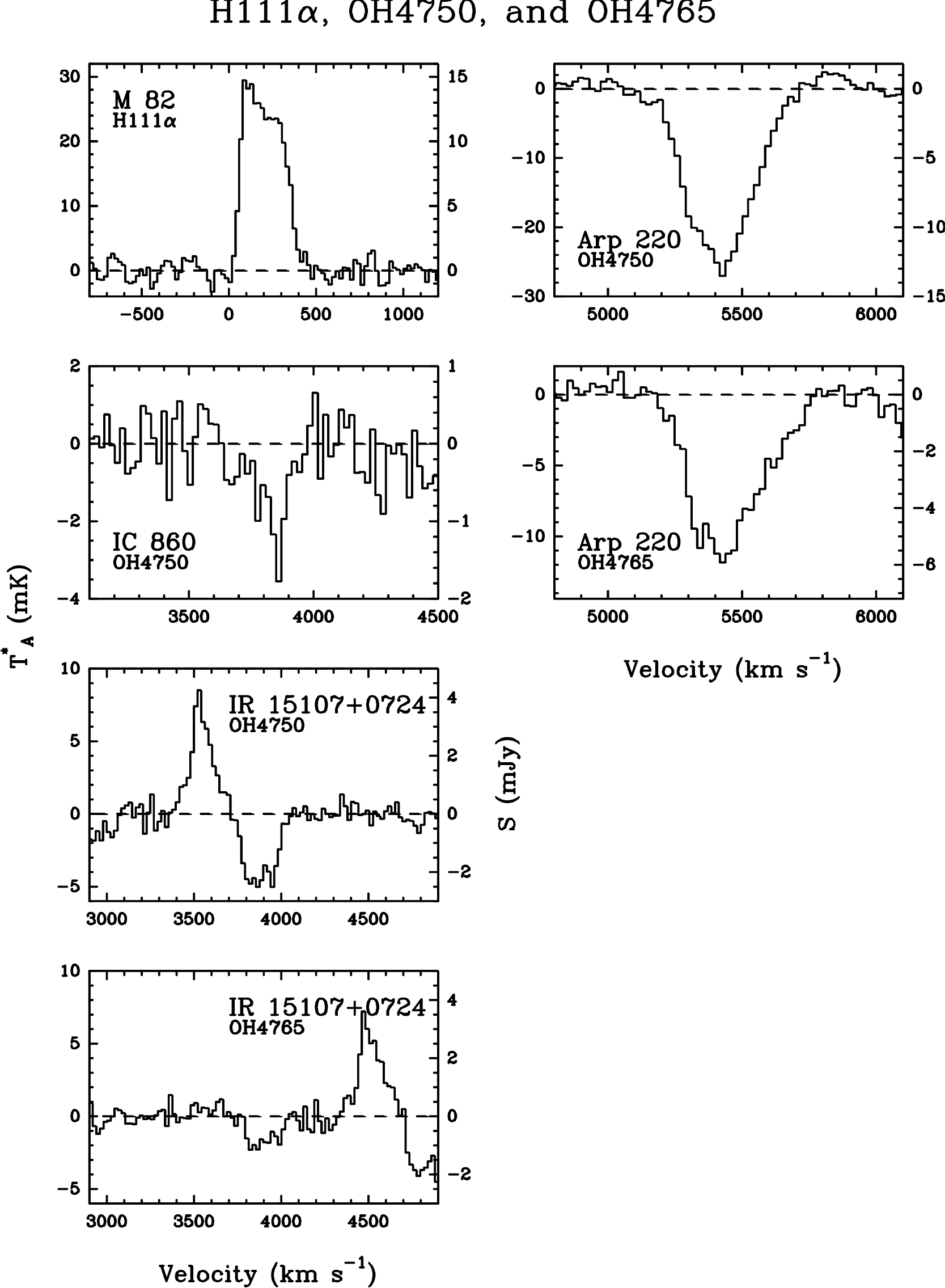}}
\caption{H111$\alpha$, OH4750 MHz, and OH4765 MHz spectra.}
\label{fig:ExgalH111aOH}
\end{figure}

Table~\ref{tab:ohh111results} lists and Figure~\ref{fig:ExgalH111aOH}
shows our measured OH $^2\Pi_{1/2} J=1/2$ $F=1-0$ and $1-1$ hyperfine
and H111$\alpha$ radio recombination line results.  For most of our
measurements we did not detect the OH or H111$\alpha$ 
transitions.  The RMS for our OH 4765 and 4750 MHz and
H111$\alpha$ measurements of the undetected sources are listed in
Table~\ref{tab:nondetections}.

Rotationally excited OH has been used as a tracer of the molecular
environment within AGN (\cite{Henkel1986}, \cite{Henkel1987},
\cite{Henkel1990}, \cite{Impellizzeri2006}).  AGN come in two main
types: those with (type 1) and those without (type 2) broad optical
line emission. In the unified scheme of active galactic nuclei, all
AGN are intrinsically similar.  A significant column density of
molecular material, in the form of a parsec-scale torus, obscures our
view of the AGN in type 2 objects.  Several surveys have attempted to
detect the obscuring molecular material using measurements of
molecular absorption or emission, yielding few confirmations 
(\eg\ \cite{Schmelz1986}, \cite{Baan1992},
\cite{StaveleySmith1992}).  On the other hand, the existence of AGN
tori are confirmed by H$_2$O megamaser emission in these objects (\cf\
\cite{Lo2005}).  The observed OH line widths for the three
sources detected in our sample are similar, though slightly larger,
than that measured in the H$_2$CO measurements of each source,
suggesting a similar dynamical origin for the OH and H$_2$CO emitting
regions.  Apparent optical depths for each of the OH and H$_2$CO line
region measurements range from 0.04 (Arp~220) to 0.22 (IC~860),
suggesting optically thin thermal absorption.

\begin{deluxetable*}{lllrrrr|rr} 
\tabletypesize{\scriptsize}
\tablewidth{0pt}
\tablecolumns{6}
\tablecaption{OH and H111$\alpha$ Detections}
\tablehead{
\colhead{Source} & 
\colhead{Transition} & 
\colhead{$T^*_A$/S$_{peak}$} & 
\colhead{v$_{hel}$} & 
\colhead{FWZI} & 
\colhead{$\int$T$^*_A$dv/$\int$Sdv} \\
&& \colhead{(mK/mJy)} & 
\colhead{(km s$^{-1}$)} & 
\colhead{(km s$^{-1}$)} & 
\colhead{(mK km s$^{-1}$/mJy km s$^{-1}$)} \\
}
\startdata
M 82          & H111$\alpha$ & 29.0(1.4)/14.7(0.7) & 203.2(5.0) & 519
& 7630(140)/3815(70)  \\ 
IC 860        & OH4750 & $-$3.6(0.8)/$-$1.8(0.4) & 3843.6(11.2) & 213
              & $-$378(73)/$-$192(37) \\ 
              & OH4765 & (0.77)/(0.39) &&& \\
IR 15107+0724 & OH4750 & $-$4.9(0.6)/$-$2.5(0.3) & 3880.2(13.8) & 410 &
               $-$1011(45)/$-$513(23) \\ 
              & OH4765 & $-$2.2(0.6)/$-$1.1(0.3) & 3891.6(13.0) & 351 &
               $-$380(45)/$-$193(23)  \\ 
Arp 220       & OH4750 & $-$28.4(1.2)/$-$14.4(0.6) & 5407.3(3.1) & 486 &
               $-$6887(122)/$-$3496(62)  \\ 
              & OH4765 & $-$13.4(1.0)/$-$6.8(0.5) & 5449.9(4.5) & 575 &
               $-$3849(110)/$-$1954(56)  \\ 
\enddata
\tablecomments{Heliocentric optical velocity frame.  RMS noise levels
  are for ~20 km~s$^{-1}$ channels.}
\label{tab:ohh111results}
\end{deluxetable*} 

\begin{deluxetable*}{cccccc} 
\tabletypesize{\scriptsize}
\tablewidth{0pt}
\tablecolumns{4}
\tablecaption{OH and H111$\alpha$ Non-Detections}
\tablehead{
\colhead{Source} & \multicolumn{3}{c}{RMS\tablenotemark{a,b}} \\
\cline{2-4}
& \colhead{OH 4765} & \colhead{OH 4750} 
& \colhead{H111$\alpha$} \\
& \colhead{(mK/mJy)} & \colhead{(mK/mJy)} & \colhead{(mK/mJy)}}
\startdata
NGC 253       & 2.22/1.11 & 4.96/2.48 & 5.56/2.78 \\
NGC 520       & 0.76/0.38 & 0.78/0.39 & 0.78/0.39 \\
NGC 660       & 0.88/0.44 & 0.96/0.48 & 0.78/0.39 \\
NGC 891       & 0.92/0.46 & 1.04/0.52 & 0.84/0.42 \\
IC 342        & 1.32/0.66 & 1.28/0.64 & 1.20/0.60 \\
Arp 55        & 1.56/0.78 & 1.62/0.81 & 1.62/0.81 \\
NGC 2903      & 1.12/0.56 & 1.24/0.62 & 1.24/0.62 \\
UGC 5101      & 1.12/0.56 & 1.08/0.54 & 1.08/0.54 \\
M 82          & 1.64/0.86 & 1.46/0.73 & Emis \\
NGC 3079      & 1.46/0.73 & 1.32/0.66 & 1.34/0.62 \\
IR 10173+0828 & 1.48/0.74 & 1.82/0.91 & 1.80/0.90 \\
NGC 3628      & 1.06/0.53 & 0.92/0.46 & 0.82/0.41 \\
M 83          & 0.78/0.39 & 0.68/0.34 & 0.86/0.43 \\
IC860         & 0.78/0.39 & Abs & 0.96/0.48 \\
IR 15107+0724 & Abs & Abs & 0.52/0.26 \\
Arp 220       & Abs & Abs & 1.60/0.80 \\
NGC 6240      & 0.72/0.36 & 0.72/0.36 & 0.84/0.42 \\
IR 17468+1320 & 1.28/0.64 & 1.60/0.80 & 1.10/0.55 \\
NGC 6946      & 0.60/0.30 & 0.58/0.29 & 0.74/0.37 \\
\enddata
\tablenotetext{a}{RMS noise levels are for 20 km~s$^{-1}$ channels.} 
\tablenotetext{b}{An ``Abs'' entry indicates that the line was detected in 
absorption, ``Emis'' in emission.} 
\label{tab:nondetections}
\end{deluxetable*} 

\subsection{Continuum Emission}
\label{ContResults}

Table~\ref{tab:contmeas} lists our measured band-center continuum
levels derived from the zero-level offset of our spectroscopic
measurements.  For most sources our measured continuum fluxes agree
with those quoted in the NED archive.  Exceptions to this agreement
are IC~342: S$_{4.8}$ = 0.087 Jy (GBT) and 0.12,0.28 (NED), M~82:
S$_{14.5}$ = 1.220/1.325 Jy (GBT) and 1.79 Jy (140$^\prime$) or 1.7 Jy
(NED), M~83: S$_{4.8}$ = 0.347 Jy (GBT) and 0.7 Jy (NED), IC860:
S$_{4.8}$ = 0.011 Jy (GBT) and 0.040 Jy (NED), and IR17468+1320:
S$_{4.8}$ = 0.208 Jy (GBT) and 0.047 Jy (NED).  For our optical depth
calculations, we use our 
measured GBT flux densities and assume an uncertainty of 15\%.
Continuum emission from starburst galaxies at these frequencies is
likely due to synchrotron emission (S$_\nu \propto \nu^{-\alpha}$)
with spectral index $\alpha$ between 0.7 and 0.8.

\begin{deluxetable*}{llll|lll} 
\tabletypesize{\scriptsize}
\tablewidth{0pt}
\tablecolumns{7}
\tablecaption{Measured Continuum Levels at Band Centers}
\tablehead{
\colhead{Source} & \multicolumn{3}{c}{4.8~GHz}& \multicolumn{3}{|c}{14.5 GHz}\\
\cline{2-4}  \cline{5-7}
& \colhead{GBT} & \colhead{Arecibo\tablenotemark{a}} & 
\colhead{NED} & \colhead{GBT} & \colhead{140$^\prime$\tablenotemark{b}} 
& \colhead{NED} \\
& \colhead{(Jy)} & \colhead{(Jy)} & \colhead{(Jy)} & \colhead{(Jy)} & 
\colhead{(Jy)} & \colhead{(Jy)}
}
\startdata
NGC 253       & 1.795 &\nodata&2.0,2.5&\nodata& 0.37 &\nodata \\
NGC 520       & 0.083 & 0.079 & 0.1   & 0.033 &\nodata&\nodata\\
NGC 660       & 0.153 & 0.140 & 0.19  & 0.059 &\nodata& $<0.09$ \\
NGC 891       & 0.324 &\nodata&0.252,0.342&\nodata&\nodata&\nodata \\
IC 342        & 0.087 &\nodata&0.12,0.28&0.045&\nodata&\nodata\\
Arp 55        & 0.039 &\nodata&\nodata&\nodata&\nodata&\nodata\\
NGC 2903      & 0.116 & 0.041 & 0.12  &\nodata&\nodata&\nodata\\
UGC 5101      & 0.061 &\nodata& 0.078 & 0.032 &\nodata&\nodata\\
M 82          & 3.266 &\nodata& 3.9   &1.220/1.325& 1.79 & 1.7\\
NGC 3079      & 0.393 &\nodata& 0.32  &\nodata& 0.14  &\nodata\\
IRAS 10173+0828&$-$0.005& 0.020&\nodata&\nodata&\nodata&\nodata \\
NGC 3628      & 0.189 & 0.131 & 0.28  &\nodata& 0.11  &\nodata\\
M 83          & 0.326 &\nodata& 0.7   & 0.075 &\nodata&\nodata\\
IC 860        & 0.010 & 0.016 & 0.040 &\nodata&\nodata&\nodata\\
IRAS 15107+0724& 0.019 & 0.031 &\nodata& 0.022 &\nodata&\nodata\\
Arp 220        & 0.213 & 0.172 & 0.2   & 0.087 & 0.11  &\nodata\\
NGC 6240       & 0.144 &\nodata& 0.178 &\nodata&\nodata&\nodata\\
IRAS 17468+1320& 0.195 &\nodata& 0.047 &\nodata&\nodata&\nodata\\
NGC 6946       & 0.170 &\nodata&0.2,0.5& 0.034 &\nodata&\nodata\\
\enddata
\tablenotetext{a}{The Arecibo 4.8~GHz continuum levels were obtained 
from spectral baselines by \citet{Araya2004}, Table~\ref{tab:sources}.}
\tablenotetext{b}{The NRAO 140$^\prime$ 14.5~GHz continuum levels were obtained 
from spectral baselines by \citet{Baan1990}, Table~\ref{tab:sources}.}
\label{tab:contmeas}
\end{deluxetable*} 

\section{Analysis}
\label{Analysis}

\subsection{Comparison to Previous Measurements}
\label{Comparison}

\noindent
{\bf NGC~253:}  Previously detected in absorption in the 4.8 and 14.5~GHz
H$_2$CO transitions \citep{Gardner1974,Baan1990}, but our GBT
detection of the $1_{10}-1_{11}$ transition is significantly different
in shape from the line profile reported by \citet{Gardner1974}.  The
interferometric measurements of the $1_{10}-1_{11}$ transition by
\citet{Baan1997} revealed an absorption structure with FWHM size
$39\arcsec\times12\arcsec$ in extent.  The peak integrated flux measured by 
\citet{Baan1997} of 1.72 Jy beam$^{-1}$ km~s$^{-1}$ is about 60\% of the
integrated flux we measure, suggesting that some extended structure
detected in our GBT observations is missing in these interferometric
measurements.  Previous CS, H$_2$CO, and high-J CO measurements
\citep{Baan1990,Baan1997,Huettemeister1997,Guesten2006} estimate
spatial densities in the range $10^{4-5}$ cm$^{-3}$.

\noindent
{\bf NGC~520:}  \citet{Araya2004} first detected the $1_{10}-1_{11}$ line
in absorption, in good agreement with the GBT line properties.  
\citet{Araya2004} detected H110$\alpha$ in NGC~520 at the 0.3~mJy level, but
our GBT observations of H111$\alpha$ lack the sensitivity to 
detect a line of similar strength (0.25~mJy rms in 40~km~s$^{-1}$
channels). Interferometric ($\theta_b = 6^{\prime\prime}$) measurements
of the CO J=$1\rightarrow0$ emission from NGC~520 \citep{Sanders1988}
revealed high concentrations of dense gas over a $12^{\prime\prime}$
(1.8 kpc) region.  \cite{Yun2001} also imaged the CO J=$1\rightarrow0$
emission from NGC~520 but with higher spatial resolution ($\theta_b =
2^{\prime\prime}.4\times2^{\prime\prime}.7$), measuring dense gas over
a somewhat smaller $7^{\prime\prime}.1\times2^{\prime\prime}.9$ region.

\noindent
{\bf NGC~660:}  Our GBT detection of the $1_{10}-1_{11}$ line in
absorption is marginal, but this line has previously been clearly
detected by \citet{Baan1986} and others.  \cite{Aalto1995} measured
the $^{12}$CO J=$1\rightarrow0$, J=$2\rightarrow1$; $^{13}$CO
J=$1\rightarrow0$, J=$2\rightarrow1$; C$^{18}$O J=$1\rightarrow0$; and
HCN J=$1\rightarrow0$ emission from this galaxy and derived a mean
source diameter of $14^{\prime\prime}$ from their CO J=$1\rightarrow0$
observations.

\noindent
{\bf NGC~891:}  This is a new detection of the $1_{10}-1_{11}$ line in 
absorption.  \cite{GarciaBurillo1995} measured the CO
J=$2\rightarrow1$ and $1\rightarrow0$ emission toward this edge-on
galaxy.  The emission is extended along the galactic disk of extent
$\gtrsim 120^{\prime\prime}\times20^{\prime\prime}$, but is mainly
concentrated in a nuclear condensation $\sim
20^{\prime\prime}\times10^{\prime\prime}$ in extent.

\noindent
{\bf IC~342:}  This is a new detection of H$_2$CO:  both lines are detected
in absorption and show similar optical depth.  Previous
millimeter-wavelength H$_2$CO measurements \citep{Huettemeister1997}
estimate spatial densities in the range $10^{4-6}$ cm$^{-3}$.
\cite{Downes1992} imaged the HCN J=$1\rightarrow0$ emission toward
this galaxy, finding that the emission is composed of five
condensations with sizes ranging from
$3.0^{\prime\prime}\times3.2^{\prime\prime}$ to
$6.5^{\prime\prime}\times3.1^{\prime\prime}$.  The total extent of
these condensations is $\sim 10^{\prime\prime}\times15^{\prime\prime}$.

\noindent
{\bf Arp~55:}  We did not detect H$_2$CO, OH, nor H111$\alpha$
emission in this galaxy.  \cite{Sanders1988} imaged the CO
J=$1\rightarrow0$ emission from this galaxy, detecting emission over a
nuclear structure $<4^{\prime\prime}$ ($<2.8$ kpc) in extent.

\noindent
{\bf NGC~2903:}  We did not detect H$_2$CO, OH, nor H111$\alpha$
emission in this galaxy.  \cite{Helfer2003} imaged the CO
$2\rightarrow1$ emission from this galaxy.  The CO emission emanates
from a nuclear structure $\sim
5^{\prime\prime}\times10^{\prime\prime}$ in extent, and from a more
extended component $\sim3^{\prime}\times10^{\prime\prime}$.

\noindent
{\bf UGC~5101:}  This is a new detection of the $1_{10}-1_{11}$ line in 
emission, and is the most distant H$_2$CO emitter known, given the 
nondetections by \citet{Araya2004} of previously claimed detections of
H$_2$CO in IRAS~12112+0305 and IRAS~10173+0828 \citep{Baan1993}.
UGC~5101 is also an H$_2$O megamaser galaxy \citep{Zhang2006}.
\cite{Imanishi2006} imaged the HCN and HCO$^+$ J=$1\rightarrow0$
emission from this ULIRG, measuring a compact emission structure
$\sim5^{\prime\prime}\times5^{\prime\prime}$ in extent.

\noindent
{\bf M~82:}  H$_2$CO was previously detected in absorption in the 
$1_{10}-1_{11}$ line by \citet{Graham1978} and in emission in the 
$2_{11}-2_{12}$ and $3_{03}-2_{02}$ (218~GHz) lines by \citet{Baan1990}.  
We detect both the 4.8 and 14.5 GHz lines in absorption with high
confidence (Figure \ref{fig:M82FormSpec}).  It is difficult to concoct
a physical scenario that would produce 14.5~GHz emission at roughly
triple the optical depth of 4.8~GHz absorption as seen by
\citet{Baan1990}, but our absorption line ratios are wholly compatible
with LVG models (see \S\ref{LVG}). The H110$\alpha$ and H111$\alpha$
lines, as well as other higher frequency hydrogen radio recombination
lines, have been detected in M~82
\citep{Seaquist1981,Bell1984,RodriguezRico2004}.  We re-detect the
H111$\alpha$ line and confirm the two components identified by
\citet{Bell1984}.  Previous CS and H$_2$CO measurements
\citep{Baan1990,Huettemeister1997} estimate spatial densities in the
range $10^{4-5}$ cm$^{-3}$.  \cite{Seaquist2006} imaged the CO
J=$6\rightarrow5$ emission from this galaxy, finding emission extended
over a disk-like structure $40^{\prime\prime}\times15^{\prime\prime}$
in extent.

\noindent
{\bf NGC~3079:}  We do not detect the $1_{10}-1_{11}$ line detected in 
absorption by \citet{Baan1986} at $-$2.1~mJy; this line would appear
as a $3.2\sigma$ feature in the GBT spectrum, but no feature below
$-$1.3~mJy ($2.0\sigma$) is seen.  Smoothing the spectrum to
$\sim100$~km~s$^{-1}$ (the line width reported by \citet{Baan1986} is
275~km~s$^{-1}$) yields a 0.4~mJy RMS noise level and no feature below
0.5~mJy at the expected line velocity.  \cite{Stevens2000} imaged the
450$\mu$m and 850$\mu$m emission from this galaxy, finding an
unresolved core $\sim20\arcsec$ in extent.  A similar source size was
measured by \cite{Braine1997} in the 1.2mm continuum emission
($\theta_s \sim 17\arcsec$).

\noindent
{\bf IRAS~10173+0828:}  \citet{Baan1993} claim a detection of the 
H$_2$CO $1_{10}-1_{11}$ line in emission, but \citet{Araya2004} do not 
detect the line at lower noise levels, and we confirm the non-detection (but 
the GBT spectrum would only detect the \citet{Baan1993} line at the $2.1\sigma$ 
level in our smoothed 20 km~s$^{-1}$ resolution spectra).
\cite{Planesas1991} imaged the CO J=$1\rightarrow0$ emission from this
galaxy, detecting an unresolved nuclear source $<6^{\prime\prime}$
($<6$ kpc) in extent.

\noindent
{\bf NGC~3628:} The $1_{10}-1_{11}$ line was detected in absorption by 
\citet{Baan1986} and confirmed by \citet{Araya2004}, while the $2_{11}-2_{12}$
line was not detected by \citet{Baan1990}.  The GBT $1_{10}-1_{11}$ line 
properties show good agreement with those measured by \citet{Araya2004}
to within the calibration uncertainties.  \cite{Stevens2005} imaged
the 450$\mu$m and 850$\mu$m continuum emission from this galaxy and
derived a core structure $\sim30^{\prime\prime}$ in size.

\noindent
{\bf M~83:}  This is a new detection of H$_2$CO:  both lines are detected
in absorption, and the 14.5~GHz line shows a significantly larger optical 
depth than the 4.8~GHz line.  \cite{Sakamoto2004} imaged the CO
J=$2\rightarrow1$ and $3\rightarrow2$ emission from this barred
starburst galaxy, deriving a
$45^{\prime\prime}\times15^{\prime\prime}$ structure.

\noindent
{\bf IC~860:}  The $1_{10}-1_{11}$ emission line was detected by 
\citet{Baan1993}, and the GBT detection shows good agreement in line
properties (and somewhat worse agreement with the Arecibo observations 
of \cite{Araya2004}).  We are not aware of any imaging measurements of
this galaxy.  \cite{Yao2003} measured the CO J=$2\rightarrow1$ and
$1\rightarrow0$ emission in $15^{\prime\prime}$ beams, which we take
as a maximum source size.

\noindent
{\bf IRAS~15107+0724:}  \citet{Baan1993} detected the $1_{10}-1_{11}$ 
line in emission, and the GBT line properties are in fair agreement
with the discovery detection and in good agreement with \citet{Araya2004}.  
We also detect the $2_{11}-2_{12}$ line in absorption.
\cite{Planesas1991} imaged the CO J=$1\rightarrow0$ emission from this
galaxy, deriving a compact nuclear source with size
$\sim3^{\prime\prime}$. 

\noindent
{\bf Arp~220/IC~4553:} The $1_{10}-1_{11}$ line of H$_2$CO was detected in 
emission by \citet{Baan1986}, but the $2_{11}-2_{12}$ line was not subsequently
detected \citep{Baan1990}.  The sub-arcsecond resolution
interferometric measurements of the $1_{10}-1_{11}$ transition by
\cite{Baan1995} detected emission from two main components with
individual sizes of $\sim 0.5^{\prime\prime}$, $\sim 1^{\prime\prime}$ in
total extent, and integrated intensities of 83 and 682
mJy beam$^{-1}$ km~s$^{-1}$.  The sum integrated intensity from these two
components compares quite favorably to the total $1_{10}-1_{11}$
integrated intensity and line width from our GBT measurements
($581\pm33$ mJy km~s$^{-1}$ over a total line width of about 500
km~s$^{-1}$).  We 
also find good agreement between our 4.8 and 14.5~GHz GBT spectra and
those reported \citet{Araya2004}.  Recent CS, HCN, and HCO$^+$
measurements \citep{Greve2007} estimate spatial densities in the
range $10^{4-6}$ cm$^{-3}$.

\noindent{The} compendium of CO, HCN, HNC, HNCO, CS, and HCO$^+$
measurements of Arp~220 presented by \citet{Greve2007} point to an apparent
dichotomy in the relative strengths of the two velocity components
exhibited by the spectra in these molecules.  The two velocity
components, at $\sim5350$ km~s$^{-1}$ (western) and $\sim5500$
km~s$^{-1}$ (eastern) have been 
associated with two nuclei embedded within a circumnuclear ring or
disk \citep{Scoville1997,Downes1998}.  \citet{Greve2007}
point out that the western velocity component is stronger
in the $^{12}$CO J=$2\rightarrow1$ and HCN J=$1\rightarrow0$
transitions, but weaker in the HCN J=$3\rightarrow2$ and
$4\rightarrow3$ transitions, than the eastern velocity component.  In
contrast, the eastern component is stronger in CS 
J=$7\rightarrow6$ and HCO$^+$ J=$4\rightarrow3$ than the western
component.  \citet{Greve2007} suggest that this dichotomy
reflects differing spatial densities in these two components: the
western component possessing lower spatial
densities, the eastern component possessing higher
spatial densities.  Our H$_2$CO measurements do not necessarily
reflect this trend.  The $1_{10}-1_{11}$ transition clearly
shows a double-horned structure, but the $2_{11}-2_{12}$
transition does not (Figure~\ref{fig:Arp220FormSpec}).

\noindent
{\bf NGC~6240:}  \citet{Baan1993} detected the $1_{10}-1_{11}$ line
in emission, but we detect the line in absorption.  It is possible that
spectral standing waves produced a spurious emission line detection in
the \citet{Baan1993} measurements.  The GBT absorption line is
detected with high confidence (5.2$\sigma$ in a single smoothed
20 km~s$^{-1}$ channel).
\citet{Greve2007} noted that $^{12}$CO, $^{13}$CO, HCN, CS, and
HCO$^+$ spectra all have a single gaussian profile which peaks at about
the same velocity of 7160 km~s$^{-1}$.  Our $1_{10}-1_{11}$ spectrum peaks
near 7300 km~s$^{-1}$, slightly blueshifted relative to the systemic velocity
of 7359 km~s$^{-1}$.  \cite{Bryant1999} imaged the CO J=$1\rightarrow0$
emission from this galaxy and derived a source size of
$10^{\prime\prime}$.

\noindent
{\bf IRAS~17468+1320:}  \citet{Baan1993} claim a tentative detection of the
H$_2$CO $1_{10}-1_{11}$ line in absorption ($3\sigma$).  The GBT observations, 
with similar spectral resolution, show no line, which would only appear at the
$2.5\sigma$ level in our smoothed 20 km~s$^{-1}$ spectra.  We are not aware
of any molecular spectral line or continuum emission imaging
measurements of this galaxy.  We arbitrarily assume a source size of
$<5^{\prime\prime}$ in our analysis.

\noindent
{\bf NGC~6946:}  This is a new detection of H$_2$CO:  the $1_{10}-1_{11}$ 
line appears in absorption.  \cite{Schinnerer2006} and
\cite{Schinnerer2007} imaged the CO J=$1\rightarrow0$,
J=$2\rightarrow1$, and HCN J=$1\rightarrow0$ emission from this
galaxy, detecting a compact nuclear source size of
$\sim2^{\prime\prime}$ and a ``nuclear spiral'' structure
$5^{\prime\prime}\times10^{\prime\prime}$ in size.

\subsection{H$_2$CO Apparent Optical Depth Calculations}
\label{OpticalDepth}

The apparent peak optical depths ($\tau$, see Equation~\ref{eq:tl})
for our H$_2$CO absorption 
measurements (Table~\ref{tab:h2codetections}) have been calculated
using the GBT continuum emission intensities listed in
Table~\ref{tab:contmeas} and assuming T$_{ex}$ = 0.  Cases where there
are large discrepancies between the GBT continuum and other
measurements are indicated by an asterisk in
Table~\ref{tab:h2codetections}.  We used the Arecibo continuum
measurement for IRAS~10173+0828 (no continuum was detected by the
GBT). Uncertainties in $\tau$ reflect only the statistical error in
the line measurement and neglect systematic errors such as calibration
uncertainties, but these should generally cancel out because the
continuum and line measurements were obtained from the same spectrum.
The optical depth listed for H$_2$CO non-detections in
Table~\ref{tab:h2codetections} is a $1\sigma$ noise level.  No optical
depth is calculated for emission lines.

\subsection{Spatial Density and Column Density Derivation Using LVG Models}
\label{LVG}

To derive the H$_2$ spatial density (number density) and H$_2$CO
column density of the dense gas in our galaxy sample, we use a model
which incorporates the Large 
Velocity Gradient (LVG) approximation (\cite{Sobolev1960}) to the
radiative transfer.  The detailed properties of our implementation of
the LVG approximation are described in \cite{Mangum1993}.  This
simplified solution to the radiative transfer equation allows for a
calculation of the global dense gas properties in a range of
environments.  As noted by \cite{Mangum1993}, one of the major sources
of uncertainty in an LVG model prediction of the physical conditions
is the uncertainty associated with the collisional excitation rates
used.  \cite{Green1991} suggests that the total collisional excitation
rate for a given H$_2$CO transition is accurate to $\sim 20\%$.
Therefore, all physical conditions predicted by our LVG model are
limited to an accuracy of no better than 20\%.

For the five galaxies where both the H$_2$CO $1_{10}-1_{11}$ and
$2_{11}-2_{12}$ transitions were detected we can derive a unique
solution to the spatial density for an assumed gas kinetic
temperature.  By fitting the intercept between the 
H$_2$CO $1_{10}-1_{11}$ and $2_{11}-2_{12}$ transition ratio and the
H$_2$CO $1_{10}-1_{11}$ transition intensity (at an assumed gas
kinetic temperature), one can derive a unique solution to the
(ensemble average) spatial 
density and H$_2$CO column density, respectively, for each galaxy.
For absorption line measurements, the LVG model fitting procedure can
also be done by using the measured apparent optical depths.  For our
galaxy sample, where $T_c \lesssim T_{cmb}$ (with one exception; M~82),
this method has the disadvantage of involving an estimate of the
excitation temperature (T$_{ex}$) in order to derive the apparent
optical depth (see \S\ref{OpticalDepth}).  Since T$_{ex}$ is dependent
upon the input physical conditions, an accurate estimate of the
apparent H$_2$CO line optical depth is not possible without an
assumption regarding the physical conditions in the gas being
modelled.  Therefore, we do not believe that fitting measured apparent
H$_2$CO optical depth to those predicted by our LVG model is a better
approach than fitting to measured H$_2$CO line temperatures.  In fact,
as we show in \S\ref{Background}, these two approaches to estimating the
excitation of H$_2$CO absorption lines in the presence of the cosmic
background radiation and weak ambient continuum emission are roughly
equivalent for the measurements presented in this paper.

A model grid of predicted H$_2$CO transition intensities over a range
in spatial density, ortho-H$_2$CO column density per velocity
gradient, and kinetic temperature of $10^{4.5}-10^{6.5}$ cm$^{-3}$,
$10^{10.6}-10^{13.8}$ cm$^{-2}$/km s$^{-1}$, and 20--70 K,
respectively, was calculated and compared to our H$_2$CO
$1_{10}-1_{11}$ and $2_{11}-2_{12}$ transition intensities
(Table~\ref{tab:h2codetections}).  Table~\ref{tab:lvg} lists the
derived LVG model best-fit physical conditions for all of the galaxies
in our sample assuming T$_{kinetic}$ = T$_{dust}$, with T$_{dust}$
from Table~\ref{tab:sources}, T$_{ex}$ = 2.73 K, and negligible
contribution due to any background continuum emission (T$_c$ = 0).  In
\S\ref{Background} we investigate how our results are affected by the
addition of a background continuum source.

\begin{deluxetable*}{lccc} 
\tabletypesize{\scriptsize}
\tablewidth{0pt}
\tablecolumns{4}
\tablecaption{Derived Physical Conditions}
\tablehead{
\colhead{Source} & 
\colhead{log(N(ortho-H$_2$CO)/$\Delta$v)\tablenotemark{a}} & 
\colhead{log(n(H$_2$))} & 
\colhead{T$_{dust}$\tablenotemark{b}} \\
& \colhead{(cm$^{-2}$/km s$^1$)} & 
\colhead{(cm$^{-3}$)} & 
\colhead{(K)}
}
\startdata
NGC~253       & $11.98\pm0.02$ & \nodata & 34 \\
NGC~520       & $11.11\pm0.07$ & \nodata & 38 \\
NGC~660       & $10.88\pm0.10$ & \nodata & 37 \\
NGC~891       & $10.89\pm0.10$ & \nodata & 28 \\
IC~342        & $11.30\pm0.15$     & $5.05\pm0.06$   & 30 \\
Arp~55        & $<11.20$ & \nodata & 36 \\
NGC~2903      & $<10.90$ & \nodata & 29 \\
UGC~5101      & $12.30\pm0.10$ & \nodata & 36 \\
M~82          & $11.60\pm0.05$     & $4.69\pm0.03$ & 45 \\
M~82 (T$_c$ = 6.6 K) & $11.08\pm0.05$     & $4.90\pm0.03$ & 45 \\
NGC~3079      & $<11.06$ & \nodata & 32 \\
IR~10173+0828 & $<11.08$ & \nodata & \nodata \\
NGC~3628      & $11.35\pm0.04$ & \nodata & 30 \\
M~83          & $11.45\pm0.24$     & $5.37\pm0.06$   & 31 \\
IC~860        & $12.40\pm0.10$ & \nodata & 40\tablenotemark{c} \\
IR~15107+0724 & $12.25\pm0.07$     & $5.68\pm0.01$ & \nodata \\
Arp~220       & $12.48\pm0.01$     & $5.66\pm0.01$ & 44 \\
NGC~6240      & $10.88\pm0.10$ & \nodata & 41 \\
IR~17468+1320 & $<10.95$ & \nodata & \nodata \\
NGC~6946      & $10.86\pm0.10$ & \nodata & 30 \\
\enddata
\tablenotetext{a}{For $1_{10}-1_{11}$ line only sources
  column density derived assuming n(H$_2$) = $10^5$ cm$^{-3}$ (for
  absorption sources) or $10^{5.65}$ cm$^{-3}$ (for emission sources) and
  T$_K$ = 40 K.}
\tablenotetext{b}{From Table~\ref{tab:sources}.}
\tablenotetext{c}{Assumed value.}
\label{tab:lvg}
\end{deluxetable*} 
\begin{figure*} 
\centering
\includegraphics[scale=0.70]{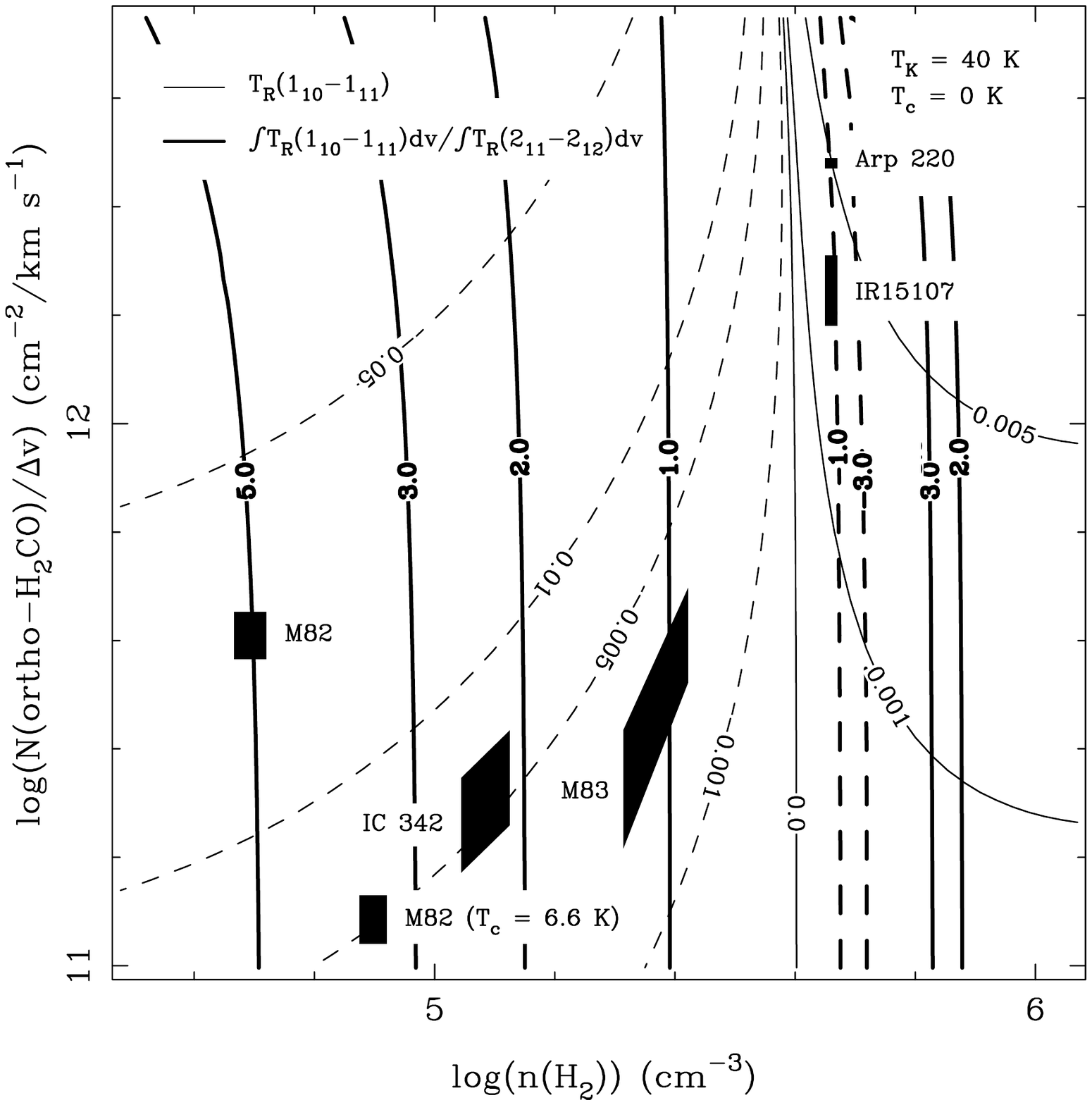}  
\caption{LVG model predictions for the spatial density (n(H$_2$)) and
  ortho-H$_2$CO column density per velocity gradient
  (N(ortho-H$_2$CO)/$\Delta$v) for the five galaxies in our sample
  with detected H$_2$CO $1_{10}-1_{11}$ and $2_{11}-2_{12}$.  Two
  model fits are shown for M~82, representing models with no continuum
  and with T$_c$ = 6.6 K at 4.83 GHz (see text).  Also
  shown are model line ratios (bold solid (positive) and dashed
  (negative) contours) and intensities (solid (positive) and dashed
  (negative) contours) for an assumed kinetic temperature of 40 K and
  no background continuum emission (T$_c$ = 0).}
\label{fig:ExgalH2COModPlot40}
\end{figure*} 

A graphical representation of our LVG model derivations of the spatial
density and ortho-H$_2$CO column density is shown in
Figure~\ref{fig:ExgalH2COModPlot40}.  Note that: 

\begin{itemize}

\item The three galaxies with H$_2$CO $1_{10}-1_{11}$ and
$2_{11}-2_{12}$ absorption (M~82, IC~342, and M~83) possess lower
column densities and generally lower spatial densities.

\item The two galaxies with H$_2$CO $1_{10}-1_{11}$ emission and
$2_{11}-2_{12}$ absorption (IR~15107+0724 and Arp~220) possess the two
largest spatial densities and column densities in our sample.

\end{itemize}

\noindent{Both} of these features of our LVG model results are due to
the unique spatial density dependence of the K-doublet transitions of
H$_2$CO.  The higher density gas in IR~15107+0724 and Arp~220 is
revealed by the H$_2$CO $1_{10}-1_{11}$ emission.  Furthermore, as
pointed out in the individual source descriptions given in
\S\ref{Comparison}, our derived spatial densities are well in-line
with estimates made based on measurements done with other
high-density tracers.

As our spatial density and ortho-H$_2$CO column density
  calculations are based on an assumed kinetic temperature, an
  understanding as to how the spatial density and ortho-H$_2$CO column
  density vary with changes in kinetic temperature is required (see
  \S\ref{LVGTkVariance}).
Following this analysis we investigate some nuances of our model fits
for M~82 (potentially extended emission and strong background
continuum) and Arp~220 (background continuum driven $1_{10}-1_{11}$
maser emission) which require some additional discussion.

\subsubsection{LVG Model Dependence on Kinetic Temperature}
\label{LVGTkVariance}

Using the same model fitting procedure described in \S\ref{LVG}, but
allowing the kinetic temperature to vary from 20 to 120 K, we can
determine the uncertainty associated with our spatial density and
ortho-H$_2$CO column density calculations due to variations in kinetic
temperature.  If the actual kinetic temperature is 120 K rather than
the assumed 40 K, for IC~342, M~82, M~83, IR~15107+0724, and Arp~220 the
spatial density decreases by 30\%, 30\%, 10\%, 50\%, and 40\%,
respectively.  Similarly, a kinetic temperature of 120 K will cause
the derived ortho-H$_2$CO column densities to rise by 100\%, 65\%,
250\%, 160\%, and 100\%, respectively.  To summarize, an
\textit{increase} by a factor of three in kinetic temperature results
in a 10--50\% \textit{decrease} in the derived spatial density and a
corresponding 65--250\% \textit{increase} in the derived ortho-H$_2$CO
column density.

\subsubsection{Influence of Possible Extended Emission in M~82}
\label{M82size}

As revealed by CO $6\rightarrow5$ (\cite{Seaquist2006},
\cite{Ward2003}), HCN $1\rightarrow0$ (\cite{Brouillet1993}), and CS
$1\rightarrow0$ and $2\rightarrow1$ \citep{Baan1990} 
images, M~82 possesses dense gas structure over an approximately
$40\arcsec\times15\arcsec$ region.  Using Equation~\ref{eq:fcoupling} we
calculate that the coupling correction for this assumed emission
distribution in our H$_2$CO $2_{11}-2_{12}$ measurements is
$f_{coupling} = 1.2$.  This correction would \textit{decrease} the measured
H$_2$CO transition ratio ($\frac{\int T_R(1_{10}-1_{11})dv}{\int
  T_R(2_{11}-2_{12})dv}$) by 20\% and produce a corresponding 
\textit{increase} in the derived spatial density.  Therefore, the
derived spatial density for M~82 is a lower-limit to the true
average spatial density in this galaxy.

\subsubsection{Influence of Background Continuum Emission}
\label{Background}

The observed line temperature assuming perfect coupling of all
continuum sources to the molecular cloud is given by

\begin{equation}
T_L = \left(T_{ex} - T_{bg}\right)\left[1-\exp(-\tau)\right]
\label{eq:tl}
\end{equation}

\noindent{where} $T_{bg}$ is the contribution due to all radiation
sources, including the 2.7 K cosmic microwave background and any
sources of continuum emission.  By over an order-of-magnitude M~82
possesses the strongest continuum emission in our sample
(Table~\ref{tab:contmeas}), thus representing an extreme case in our
analysis.  We have 
modelled the influence of a significant source of background continuum
emission in our LVG models of M~82 by including an additional
radiation source whose spectral characteristics and intensities mimic
our measured 4.8 and 14.5 GHz continuum emission.  The effect that
this additional radiation source has on our derived spatial density
and H$_2$CO column density is shown in 
Figure~\ref{fig:ExgalH2COModPlot40}.  Relative to the no-continuum
model fit, the derived spatial density and H$_2$CO column density are
$\sim 1.5$ times larger and $\sim 3$ times smaller, respectively, when
the observed 
background continuum emission is included.  This is an expected
result, as the relatively stronger continuum emission source at 4.8
GHz provides a larger contribution to the excitation of the observed
H$_2$CO $1_{10}-1_{11}$ line emission, lowering the model predicted
transition ratio ($\frac{\int T_R(1_{10}-1_{11})dv}{\int
  T_R(2_{11}-2_{12})dv}$).  Therefore, ignoring contributions due to
background continuum emission produces a derived spatial density for
M~82 which is a lower-limit to the true average spatial
density in this galaxy.

A further check of the influence of the continuum emission from our
galaxy sample in the excitation of the H$_2$CO molecule is afforded by
a comparison between the apparent H$_2$CO optical depth estimates and
those predicted by our LVG models where we have assumed $T_c = 0$.
For the three galaxies where both the $1_{10}-1_{11}$ and
$2_{11}-2_{12}$ transitions are observed to be in absorption (IC~342, M~82 with
T$_c$ = 6.6 K, M~83), line optical depths, within both the modelling and
optical depth calculation uncertainties, are consistent.  For the two
galaxies where only the $2_{11}-2_{12}$ transition is observed in absorption
(IR~15107 and Arp~220), LVG model-predicted optical depths
are a factor of two larger than the calculated apparent optical
depths.  Since the apparent optical depth calculation assumes that
$T_{ex} = 0$, our apparent optical depth estimates are likely to be
lower limits to the true line optical depth.  Given the approximations
made to derive the apparent and LVG modelled optical depths, we
suggest that there is good consistency between our continuum-less
model-predicted optical depths and those estimated by our apparent
absorption line optical depth calculations.

\subsubsection{H$_2$CO Excitation Pumping by Continuum Emission in Arp~220}
\label{Arp220Cont}

Bright and compact nonthermal continuum emission at frequencies near
4.8 GHz can induce maser excitation of the H$_2$CO
$1_{10}-1_{11}$ transition (\cite{Baan1986}).  This maser excitation
would amplify the background continuum emission and be 
observed as anomalously bright H$_2$CO (maser) emission.  Assuming an
H$_2$CO and nonthermal continuum emission source size of $\sim
0.25^{\prime\prime}$ and nonthermal continuum brightness temperature
of $\sim 2\times10^5$ K, \cite{Baan1986} derived an H$_2$CO
$1_{10}-1_{11}$ brightness temperature of $\sim
3\times10^4$ K.  To study the conditions under which such a high
brightness temperature would be produced Baan \etal\ constructed an
LVG model which included the effects of a strong nonthermal radiation
field.  These models showed that weak amplification ($\tau_{4.8GHz} >
-0.025$) of the $1_{10}-1_{11}$ transition results for
n(H$_2$) $< 10^5$ cm$^{-3}$ and H$_2$CO abundances of $>10^{-9}$.

Our measurement of H$_2$CO $2_{11}-2_{12}$
\textit{absorption} in Arp~220 puts some significant constraints on the
potential for maser amplification of the $1_{10}-1_{11}$
transition.  We have reproduced the LVG model predictions described in
\cite{Baan1986}, but also find that the conditions which lead to 
amplification of the $1_{10}-1_{11}$ transition also
predict:

\begin{itemize}
\item Weak ($\tau_{14.5GHz} \simeq 0.5\times\tau_{4.8GHz}$) maser
  amplification of the $2_{11}-2_{12}$ transition,
\item $\frac{1_{10}-1_{11}}{2_{11}-2_{12}}$
  transition intensity ratios which are not consistent with our
  measured results.
\end{itemize}

This result forces us to conclude that either: (1) the
$1_{10}-1_{11}$ transition is not masing or (2) the two
K-doublet transitions do not trace the same dense gas component.
Note that this result is independent of the assumed kinetic
  temperature (see \S\ref{LVGTkVariance}).
Without high spatial resolution measurements of the $2_{11}-2_{12}$
transition we cannot discount the non-cospatiality conclusion. 
Based on the nearly identical line profiles from 5200
to 5700 km~s$^{-1}$, we believe that the two K-doublet transitions trace the
same dense gas component.  Therefore, we conclude that the
$1_{10}-1_{11}$ emission in Arp~220 is not due to maser
emission, but does represent the signature of a dense gas component
near the two starburst nuclei in this galaxy.

\subsubsection{Single Transition H$_2$CO Column Density Estimates}
\label{Colden}

\begin{figure}
\resizebox{\hsize}{!}{
\includegraphics[scale=0.75]{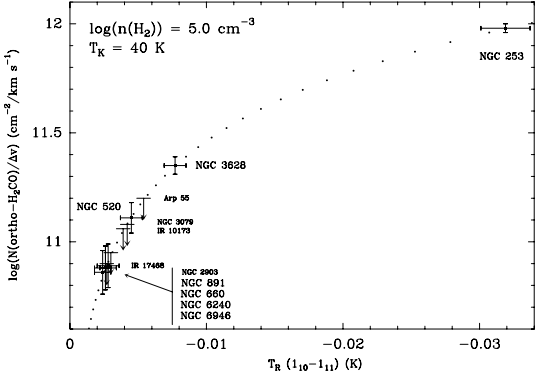}} 
\caption{LVG model predictions for the ortho-H$_2$CO column density
  per velocity gradient (N(ortho-H$_2$CO)/$\Delta$v) for the seven
  (large font) galaxies in our sample with detected H$_2$CO
  $1_{10}-1_{11}$ absorption only.  We also show upper
  limits ($3\sigma$) to N(ortho-H$_2$CO)/$\Delta$v for our five
  non-detected (small font) galaxies.  Not shown on this plot are the
  two $1_{10}-1_{11}$ emission line sources IC~860 and UGC~05101.}
\label{fig:colden}
\end{figure}

For a given spatial density and kinetic temperature a single H$_2$CO
transition can be used to derive the H$_2$CO column density.  To
estimate H$_2$CO column densities toward the galaxies that we have
only detected in the $1_{10}-1_{11}$ transition, we assume a spatial
density of $10^5$ cm$^{-3}$ (for absorption sources) or $10^{5.65}$
cm$^{-3}$ (for emission sources) and kinetic temperature of 40 K.  Our
spatial density assumption is based upon our derivations of the
spatial density in the five galaxies where both H$_2$CO transitions
were measured, while our assumed kinetic temperature is typical of our
galaxy sample (see Table~\ref{tab:sources}).  Using our measured
$1_{10}-1_{11}$ intensities we derive the H$_2$CO column densities for
each galaxy (Figure~\ref{fig:colden} and Table~\ref{tab:lvg}).

\subsection{A Cautionary Note Regarding Critical Density}
\label{CriticalDensity}

Very often estimates of the spatial density in molecular clouds are
based on a calculation of the ``critical density'' of a measured
molecular transition.  The critical density of H$_2$ (n$_{crit}$) is
given by: 

\begin{equation}
n_{crit}(cm^{-3}) = \frac{A_{ul}(s^{-1})}{\sum_iC_{ui}(cm^3~s^{-1})}
\label{eq:ncrit}
\end{equation}

\noindent{where} $A_{ul}$ is the spontaneous emission (Einstein A)
rate for the transition between levels u and l, while $C_{ui}$ is the
H$_2$ collisional excitation rate out of the upper energy level into level
i.  This critical density is really a measure of the relative
importance of radiative optically-thin and collisional excitation to
the population of the upper state in a molecular transition.  $A_{ul}$
is given by: 

\begin{eqnarray}
A_{ul}(s^{-1}) &\equiv& \frac{64 \pi^4 \nu^3}{3hc^3}|{\mathbf \mu_{lu}}|^2 \\
      &=& \frac{64 \pi^4 \nu^3}{3hc^3} S {\mathbf \mu}^2
\label{eq:adef}
\end{eqnarray}

\noindent{where} $|{\mathbf \mu_{lu}}|^2$ is the dipole moment matrix
element for the transition involving levels l and u, $\mu$ is the
dipole moment (2.331 Debye; \citet{Kondo1960}), and $S$ is the line
strength, given by:

\begin{equation}
S = \frac{K^2}{J(J+1)} \textrm{ for $(J,K)\rightarrow (J,K)$}
\label{eq:linestrengthlsq}
\end{equation}

\noindent{where} we use the (very good) approximation that H$_2$CO is
a symmetric top molecule.  For the $1_{10}-1_{11}$ and
$2_{11}-2_{12}$ transitions $S = \frac{1}{2}$ and $\frac{1}{6}$,
respectively.  Using these line strengths, along with $\mu$, in
Equation~\ref{eq:adef} yields $A_{ul} = 3.6\times10^{-9}$ and
$3.2\times10^{-8}$ s$^{-1}$, respectively.  The sum over the
collisional excitation rates\footnote{Note that H$_2$CO collisional
  excitation rates are calculated using He as collider.  The quoted
  rates have been scaled by a factor of 2.2 to correct for the
  difference between He and H$_2$ collisions \citep{Green1991}.} out
of level u ($\sum_iC_{ui}$) are $4.4\times10^{-10}$ and
$4.2\times10^{-10}$ cm$^3$~s$^{-1}$, respectively, at a kinetic temperature
of 40 K.  This implies that the 
critical densities for the $1_{10}-1_{11}$ and
$2_{11}-2_{12}$ transitions are 8 and 76 cm$^{-3}$,
respectively.  These critical densities are at least three
orders-of-magnitude lower than critical densities for low-lying
transitions of HCN, HCO$^+$, and CS (\cf\ \cite{Jansen1995},
\cite{Evans1999}) which 
have been used as high-density tracers of star formation.  As we are
clearly measuring densities which are many 
orders of magnitude larger than the critical density for the H$_2$CO
transitions studied in this work, the critical density appears to be a
poor indicator of the density probing properties of a molecular
transition.  Since significant optical depth in a transition can lower
the critical density by limiting the radiative contribution to the
upper energy level population, the critical density is not actually a
lower limit to the density required to excite a molecular transition
under many interstellar conditions.  Clearly density estimates based
on critical density arguments are nothing more than crude estimates to
the true spatial density in a molecular cloud.

\subsection{H$_2$CO Luminosity and Dense Gas Mass}
\label{H2COLuminosityandMass}

Studies of the correlation between the infrared and molecular line
luminosity in luminous infrared galaxies have been used to
characterize the power source for their intense infrared emission.
Surveys of the HCN (\citet{Gao2004b}, \citet{Gao2004b}) and HCO$^+$
(\citet{GarciaCarpio2006}) emission toward these galaxies
have derived a tight correlation between their molecular line and
infrared luminosities.  A comparison between the infrared and HCN
J=$1\rightarrow0$ emission in our Galaxy has extended this correlation
down to Galactic dense core scales, extending the correlation over
7--8 orders of magnitude in L$_{IR}$ (\citet{Wu2005}).  This
correlation has been 
interpreted as evidence that star formation is the main power source
for the large infrared luminosities observed in these galaxies.

Following the formalism for calculating molecular spectral line 
luminosity elucidated by \cite{Solomon1997}, we note that the
monochromatic luminosity $L(\nu_{rest}$), the observed flux density
$S(\nu_{obs}$), and the luminosity distance $D_L$ are related by
$\nu_{rest} L(\nu_{rest}) = 4\pi S(\nu_{obs}) \nu_{obs} D^2_L$.
Relating $L(\nu_{rest})$ to the spectral line luminosity $L_{line}$:

\begin{eqnarray}
L_{line} &=& \int L(\nu_{rest}) d\nu \nonumber \\
         &=& 4\pi S_{line}(\nu_{obs}) \nu_{obs} D^2_L \frac{dv}{c} \nonumber \\
         &=& 1.043\times10^{-3} S_{line} \Delta v \nu_{rest}
         (1+z)^{-1} D^2_L~L_\odot
\label{eq:linelumin}
\end{eqnarray}

\noindent{with} $S_{line}$ in Jy, $\Delta v$ in km~s$^{-1}$, $\nu_{rest} =
\nu_{obs}(1+z)$ in GHz, and $D_L$\footnote{$D_L = D_A\left(1+z\right)^2$, where
  $D_L$ is the luminosity distance and $D_A$ is the angular size
  distance.} is in Mpc.  As \cite{Solomon1997} point 
out, it is convenient for relating spectral line integrated intensity
measurements to introduce the quantity $L^\prime$, which is the
source-integrated surface brightness.  Two spectral lines with the
same main beam brightness temperature ($T_{mb}$) and spatial extent
will have the same $L^\prime$, making the $L^\prime$ ratio of two
spectral line measurements an indicator of the physical conditions in
the gas.

For any molecular spectral line measurement $L^\prime$ is given by:

\begin{eqnarray}
L^\prime_{line} &=& \int T_{b} dv d\Omega_s D^2_A \nonumber \\
               &=& \int T_{mb} dv \Omega_{s\ast b} (1+z) D^2_A \nonumber \\
               &=& 23.504 \Omega_{s\ast b}(arcsec^2) (1+z)^{-3}
                   D^2_L(Mpc) \nonumber \\
               && \int T_{mb}(line) dv~K~km~s^{-1}~pc^2
                   \nonumber \\
               &=& \frac{23.504~\pi\left(\theta^2_s(arcsec) +
                   \theta^2_B(arcsec)\right)}{4\ln(2)} (1+z)^{-3} \nonumber \\
               && D^2_L(Mpc) \int T_{mb}(line) dv~K~km~s^{-1}~pc^2 \\
               &=& \frac{c^2}{2k\nu^2_{rest}} (1+z)^{-1} D^2_L 
                   \int S_{line}~dv \nonumber \\ 
               &=& 3.256\times 10^7 \nu^{-2}_{rest}
               (1+z)^{-1} D^2_L \nonumber \\
               && \int S_{line}~dv~Jy~km~s^{-1}~pc^2
\label{eq:lprime}
\end{eqnarray}

\noindent{with} $S_{line}$ in Jy, $T_{mb}(line)$ in K, dv in km~s$^{-1}$,
$\nu_{rest} = \nu_{obs}(1+z)$ in GHz, $D_L$ in Mpc, and 
$\Omega_{s\ast b}$ the solid angle of a gaussian source convolved with the
gaussian telescope beam, given by:

\begin{equation}
  \Omega_{s\ast b} = \frac{\pi\theta_B^2}{4\ln(2)}
\left(\frac{\theta^2_s + \theta^2_B}{\theta^2_B}\right)
\label{eq:omegasb}
\end{equation}

Studies which compare L$_{IR}$ with L$^\prime$ are seeking to
investigate the correlation between source-integrated dense gas column
density and mass to IR luminosity.  Since we have LVG-derived H$_2$CO
column densities, the mass of dense gas traced by H$_2$CO can be
derived as follows:

\begin{eqnarray}
M_{dense} &=& N_{mol} \Omega_s D^2_A \frac{\mu m_{H_2}}{X_{mol}} \nonumber \\ 
               &=& 23.504 \Omega_s(arcsec^2) (1+z)^{-2}
                   D^2_L(Mpc) \nonumber \\
               && \frac{N_{mol}(cm^{-2}) \mu m_{H_2}(gm)}{X_{mol}}~pc^2~cm^{-2} 
                   \nonumber \\
               &=& \frac{1.125\times10^5~\pi\theta^2_s(arcsec)
                 D^2_L(Mpc)}{4\ln(2) (1+z)^{2}} \nonumber \\
               && \frac{N_{mol}(cm^{-2}) \mu m_{H_2}(gm)}{X_{mol}}~M_{\odot}
\label{eq:mdense}
\end{eqnarray}

\noindent{where} $m_{H_2}$ is the mass of molecular hydrogen, $\mu$
accounts for the mass fraction due to He, and $X_{mol}$ is the
abundance (relative to H$_2$) of the molecule.

\begin{deluxetable*}{lllll} 
\tabletypesize{\scriptsize}
\tablewidth{0pt}
\tablecolumns{5}
\tablecaption{Infrared Luminosities and Dense Gas Masses}
\tablehead{
\colhead{Source} &
\colhead{Size\tablenotemark{a}} & 
\colhead{L$_{IR}$\tablenotemark{b}} &
\colhead{M$_{dense}$(HCN)\tablenotemark{c}} &
\colhead{M$_{dense}$(H$_2$CO)} \\
& \colhead{(arcsec)} &
\colhead{($10^{10} L_\odot$)} & 
\colhead{($10^8 M_\odot$)} &
\colhead{($10^8 M_\odot$)}
}
\startdata

NGC 253        & $39\times12$ & 2.78$\pm$0.001 & 2.7 & 3.02$\pm$2.14 \\

NGC 520        & 12 & 8.10$\pm$0.05 & \nodata & 10.16$\pm$7.37 \\

NGC 660        & 14 & 3.10$\pm$0.01 & $>2.6$ & 5.84$\pm$4.34 \\

NGC 891        & $20\times10$ & 1.86$\pm$0.005 & 2.5 & 1.39$\pm$1.04 \\

IC 342         & $10\times15$ & 1.47$\pm$0.002 & 4.7 & 0.14$\pm$0.11 \\

Arp 55         & $<4$ & 45.37$\pm$1.33 & 38 & $<$56.91 \\

NGC 2903       & $\sim5\times15$ & 1.55$\pm$0.004 & $>0.9$ & $<$0.22 \\

UGC 5101       & $\sim5$ & 83.11$\pm$1.03 & 100 & 16.61$\pm$12.25 \\

M 82           & $40\times15$ & 5.92$\pm$0.001 & 3.0 & 3.81$\pm$2.73 \\	

NGC 3079       & $\sim20$ & 5.37$\pm$0.02 & $\sim10$ & $<$25.27 \\

IRAS 10173+0828 & $<6$ & 57.17$\pm$1.79 & \nodata &
$<$485.44\tablenotemark{d} \\ 

NGC 3628        & $\sim30$ & 1.78$\pm$0.008 & 2.4 & 19.56$\pm$13.95 \\

M 83            & $45\times15$ & 1.26$\pm$0.001 & 3.5 & 1.14$\pm$1.05 \\

IC 860          & $<15$ & 14.48$\pm$0.10 & \nodata & $<581.19$ \\

IRAS15107+0724 & $\sim3$ & 18.25$\pm$0.10 & \nodata & 32.22$\pm$23.37 \\

Arp 220        & $\sim1$ & 155.76$\pm$0.32 & 30--370 & 19.85$\pm$14.05 \\

NGC 6240       & 10 & 67.99$\pm$0.46 & 80--260 & 77.39$\pm$57.60 \\

IRAS 17468+1320& $<5$  & 8.71$\pm$0.18 & \nodata & $<$5.93\tablenotemark{e} \\

NGC 6946       & $5\times10$ & 1.43$\pm$0.002 & 4.9 & 0.06$\pm$0.05 \\

\enddata
\tablenotetext{a}{See \S\ref{Comparison} for references.}
\tablenotetext{b}{IRAS fluxes from \citet{Sanders2003}.}
\tablenotetext{c}{Except for Arp~220 and NGC~6240 \citep{Greve2007},
  from \cite{Gao2004a}.}
\tablenotetext{d}{Assumed H$_2$CO linewidth from \cite{Araya2004}}
\tablenotetext{e}{Assumed H$_2$CO linewidth from \cite{Baan1986}}
\label{tab:luminosity}
\end{deluxetable*} For emulateapj

In Table~\ref{tab:luminosity} we list the infrared luminosities
derived from measurements culled from the literature along with the
dense gas mass (M$_{dense}$) derived from HCN and our H$_2$CO
measurements.  The infrared luminosity $L_{IR}$ over the wavelength
range from 8 to 1000 $\mu$m is derived from
the \citet{Sanders1996} relation

\begin{eqnarray}
L_{IR} &=& 4\pi D^2_L F_{IR} \\
F_{IR} &=& 1.8\times 10^{-14} (13.48 f_{12} \nonumber \\
      && + 5.16 f_{25} + 2.58 f_{60} + f_{100}) W m^{-2}
\label{eq:lir}
\end{eqnarray}

\noindent{with} $f_{12}$, $f_{25}$, $f_{60}$, $f_{100}$ given by the
IRAS 12~$\mu$m, 25~$\mu$m, 60~$\mu$m, and 100~$\mu$m fluxes in Jy,
respectively.  The H$_2$CO-derived dense gas masses were calculated
using Equation~\ref{eq:mdense} with N(ortho-H$_2$CO)/$\Delta$v from
our LVG model fits (Table~\ref{tab:lvg}), $\Delta$v derived from
our direct-measurements (Table~\ref{tab:h2codetections}), and assuming
X(H$_2$CO) = $10^{-9}$.  In addition to the derived uncertainties in
N(ortho-H$_2$CO)/$\Delta$v and $\Delta$v, we have
  conservatively applied 50\% uncertainties each to our source sizes
  and H$_2$CO abundances when calculating $M_{dense}$.  Due to the
  lack of H$_2$CO source structure information our H$_2$CO-derived
  M$_{dense}$ values are only estimates to be compared with other
  molecule-derived dense gas masses.

Our $M_{dense}$ estimates range from $6\times10^6 M_\odot$
  (NGC~6946) to 
$8\times10^9 M_\odot$ (Arp~220).  In general, $M_{dense}$ derived
from our H$_2$CO measurements agrees to within a factor of 3 with
$M_{dense}$ derived from HCN measurements.  In a few cases, such as
IC~342, UGC~05101, NGC~3628, and NGC~6946, the dense gas mass derived
from H$_2$CO is as much as a factor of 30 too low relative to
$M_{dense}$ derived from HCN.  In the case of IC~342 we can also
calculate $M_{dense}$ using our derived n($H_2$)
(Table~\ref{tab:lvg}).  Assuming a thin-disk for the emitting volume
the mass is derived from
\begin{equation}
  M^\prime_{dense} = \frac{\pi\theta^2_s}{4\ln(2)}D^3_l h m_{H_2} n(H_2)
\label{eq:mdenseprime}
\end{equation}
where $h$ is the thickness of the galactic disk.  For IC~342 we find
that $M^\prime_{dense} \simeq 2\times10^9 M_\odot$ assuming $h$ = 20
pc ($\sim\frac{1}{5}\theta_s$).  This value for the dense gas mass is
more in-line with other estimates of $M_{dense}$ (\ie\ from HCN).  One
way to bring our estimates of $M_{dense}$ and $M^\prime_{dense}$ into
agreement is to use a lower H$_2$CO abundance of $10^{-10}$.

\cite{Gao2004a} found that $M_{dense}$(HCN) are factors of 5--200
smaller than $M(H_2)$ derived from CO measurements for their sample of
spiral, LIRG, and ULIRG galaxies, many of which are included in our
sample.  As pointed out by \cite{Gao2004a}, \cite{Gao2004b}, and
\cite{Greve2007}, the progressive difference between the mass traced
by dense gas tracers such as CO, HCN, and H$_2$CO reflects the
hierarchical structure of the giant molecular clouds in these galaxies.
Our $M_{dense}$(H$_2$CO) values suggest that H$_2$CO, like HCN, traces
a denser, more compact, component of the giant molecular clouds in our
galaxy sample than CO. This result is consistent with high-resolution
studies of the K-doublet H$_2$CO emission in our own Galaxy (\cf\
\cite{Mangum1993a}).

\begin{figure*} 
\centering
\includegraphics[angle=-90,scale=0.63]{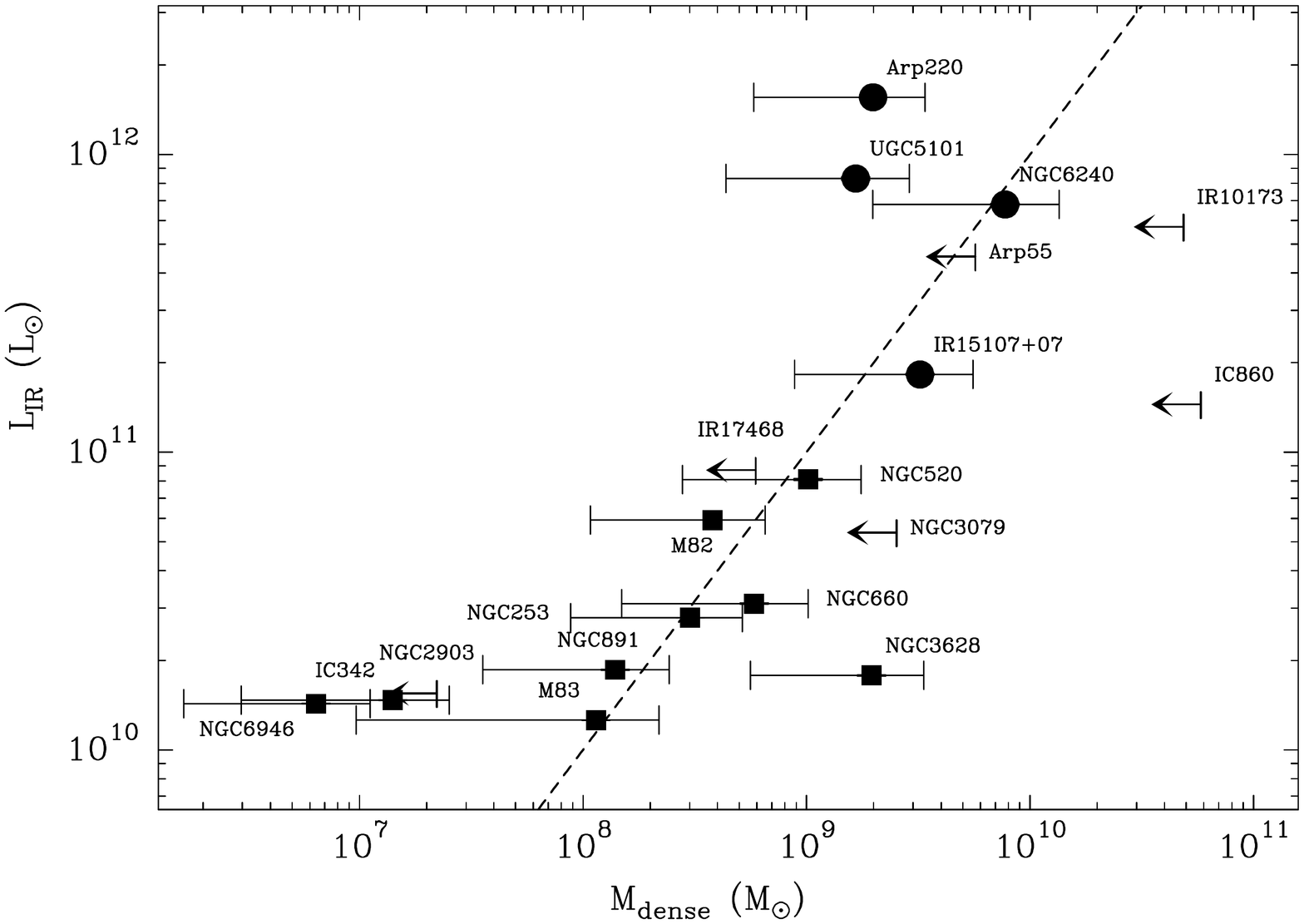}
\caption{$L_{IR}$ versus $M_{dense}$.  $M_{dense}$ has been derived
  using Equation~\ref{eq:mdense} assuming $X(H_2CO) = 10^{-9}$ and
  uncertainties calculated assuming 50\% errors in $\Omega_s$ and
  $X(H_2CO)$.  Filled circles and squares are used to indicate
  H$_2$CO-detected galaxies at distances $>$ and $\leq$ 50 Mpc,
  respectively.  The dotted line represents the log-linear relation
  $\log(L_{IR}) = 2 + 1.0 \log(M_{dense})$.}
\label{fig:lirmdense}
\end{figure*} 

In Figure~\ref{fig:lirmdense} we show $M_{dense}$ versus $L_{IR}$.
Note the trend, similar to that derived by \cite{Gao2004a} using
HCN J=$1\rightarrow0$ measurements.  We also show in
Figure~\ref{fig:lirmdense} the log-linear relation $\log(L_{IR}) =
2+1.0\log(M_{dense})$.  Note that given the uncertainties regarding
dense gas mass source size we have not performed a formal fit of this
relation.  The linear correlation between the dense gas mass
traced by H$_2$CO and the infrared luminosity is consistent with 
the correlation first noted by \cite{Solomon1992}.  Active star
formation in IR-bright galaxies is clearly driven by the amount of
material available to form stars.  Furthermore, the relationship
to Kennicutt-Schmidt laws, which relate the star formation rate to
the mass of gas available to produce stars
(\cite{Schmidt1959,Schmidt1963}) in galaxies, to these measurements
becomes apparent in the $L_{IR}$-to-$M_{dense}$ correlation.  As shown
by \cite{Krumholz2007}, the actual form for this correlation depends
upon the spectral line chosen to derive the spectral line luminosity
$L_{mol}$.  Since $M_{dense} \propto L_{mol}$, the suggestion of a
linear correlation between $L_{IR}$ and $M_{dense}(H_2CO)$ implies
that, similar to HCN, H$_2$CO traces the dense star-forming gas in
starburst galaxies.

A related trend can be seen when one compares $L_{IR}$ to our derived
$n(H_2)$ for the five galaxies within which we have derived spatial
densities.  From Tables~\ref{tab:lvg} and \ref{tab:luminosity} we note
that the three galaxies with $L_{IR} \simeq 1-5\times 10^{10} L_\odot$
have derived spatial densities ($n(H_2) \simeq 10^5 cm^{-3}$).  The
two higher luminosity galaxies $L_{IR} > 10^{11} L_\odot$ have higher
derived spatial densities of $n(H_2) = 10^{5.7} cm^{-3}$.  Even though
our number statistics are small (five galaxies), there appears to be a
trend toward higher spatial density for galaxies with higher infrared
luminosity.  This is likely another representation of the
$L_{IR}$-$M_{dense}$ correlation.

\section{Conclusions}
\label{Conclusions}

Using measurements of the $1_{10}-1_{11}$ and $2_{11}-2_{12}$
K-doublet transitions of H$_2$CO we have derived accurate
\textit{measurements} of the spatial density (n(H$_2$)) in a sample of
starburst galaxies.  The derived densities range from
$10^{4.7}$-$10^{5.7}$~cm$^{-3}$, consistent with the suggestion that the
high infrared brightness of these galaxies is driven by extreme star
formation activity.  We believe that these spatial density
measurements are the most accurate measurements of this important
physical quantity in starburst galaxies made to-date.  We have also used our
H$_2$CO measurements to derive a measure of the dense gas mass which
ranges from $0.06-77\times10^8 M_\odot$, generally consistent with
previous measurements, mainly from studies of the HCN emission in
these galaxies.  The linear correlation between the IR luminosity
($L_{IR}$) and the dense gas mass ($M_{dense}$) found in larger
samples of the HCN emission in starburst galaxies is also apparent in
our H$_2$CO measurements.  This further supports the suggestion that
active star formation in IR-bright galaxies is driven by the amount of 
material available to form stars.  We also note a related trend
between $L_{IR}$ and our derived $n(H_2)$ for the five galaxies within
which we have derived spatial densities.  The three galaxies with
lower IR luminosities have lower ($n(H_2) \simeq 10^5$~cm$^{-3}$)
derived spatial densities, while the two higher luminosity galaxies
have higher ($n(H_2) = 10^{5.7}$~cm$^{-3}$) derived spatial densities.
This is likely another representation of the $L_{IR}$-$M_{dense}$
correlation.

\acknowledgments

The GBT staff, especially Frank Ghigo, were characteristically helpful
and contributed significantly to the success of our observing program.
JGM acknowledges many fruitful discussions with Paul Vanden Bout, Al
Wootten, J\"urgen Ott, and Harvey Liszt (the ``Oracle of
Charlottesville'').  We also thank our anonymous referee for providing
several very good comments and suggestions which significantly
improved this presentation.  Support for this work was provided by NASA
through Hubble Fellowship grant \#HST-HF-01183.01-A awarded by the 
Space Telescope Science Institute, which is 
operated by the Association of Universities for Research in Astronomy, 
Incorporated, under NASA contract NAS5-26555.
This research has made use of the NASA/IPAC Extragalactic Database (NED) 
which is operated by the Jet Propulsion Laboratory, California Institute 
of Technology, under contract with the National Aeronautics and Space
Administration.

\textit{Facilities:} \facility{GBT}


\end{document}